# Illusion and Reality in the Atmospheres of Exoplanets

## L. Drake Deming[1,2], and Sara Seager[3]

[1]Department of Astronomy, University of Maryland at College Park, College Park MD, 20742

[2]NASA Astrobiology Institute's Virtual Planetary Laboratory

[3]Department of Earth and Planetary Sciences, and Department of Physics, Massachusetts Institute of Technology, Cambridge MA, 02139

Corresponding author: Drake Deming (ddeming@astro.umd.edu)

**Key Points:**

- The atmospheres of exoplanets are the window into all of their properties beyond mass, radius, and orbit.
- The field moves from the first tentative and often incorrect conclusions, converging to the reality of exoplanetary atmospheres.

**Abstract.** The atmospheres of exoplanets reveal all their properties beyond mass, radius, and orbit. Based on bulk densities, we know that exoplanets larger than 1.5 Earth radii must have gaseous envelopes, hence atmospheres. We discuss contemporary techniques for characterization of exoplanetary atmospheres. The measurements are difficult, because - even in current favorable cases - the signals can be as small as 0.001% of the host star's flux. Consequently, some early results have been illusory, and not confirmed by subsequent investigations. Prominent illusions to date include polarized scattered light, temperature inversions, and the existence of carbon planets. The field moves from the first tentative and often incorrect conclusions, converging to the reality of exoplanetary atmospheres. That reality is revealed using transits for close-in exoplanets, and direct imaging for young or massive exoplanets in distant orbits. Several atomic and molecular constituents have now been robustly detected in exoplanets as small as Neptune. In our current observations, the effects of clouds and haze[1] appear ubiquitous. Topics at the current frontier include the measurement of heavy element abundances in giant planets, detection of carbon-based molecules, measurement of atmospheric temperature profiles, definition of heat circulation efficiencies for tidally-locked planets, and the push to detect and characterize the atmospheres of super-Earths. Future observatories for this quest include the James Webb Space Telescope, and the new generation of Extremely Large Telescopes on the ground. On a more distant horizon, NASA's concepts for the HabEx and LUVOIR missions could extend the study of exoplanetary atmospheres to true twins of Earth.

## 1.0 Introduction: Why we study exoplanetary atmospheres

The study of extrasolar planets ("exoplanets") is one of the fastest growing sub-disciplines in astronomy and planetary science. In the past 25 years, we have gone from being ignorant of whether exoplanets even exist in significant numbers, to the realization that they are abundant in our Galaxy. We have learned that many exoplanets are strikingly different than the planets of our Solar System.  Most important, we now have dozens of alien atmospheres than are amenable to observations, so that we can measure their properties and begin to conduct comparative studies.

---

[1] Haze is defined as very small particles suspended in the atmosphere that are opaque, diminish visibility, and contribute to color. Hazes are almost always photochemically produced

The atmospheres of exoplanets can host biosignature gases (e.g., Des Marais et al. 2002), so the path to finding possible signs of life on exoplanets most likely leads through their atmospheres. Indeed, the atmosphere is the window into all exoplanetary properties beyond mass, radius, and orbital dynamics.  Moreover, exoplanetary atmospheres host a wide variety of fascinating physical processes, from intense infrared radiation and exotic chemistry, to super-sonic winds and electric currents. One of our best ways to probe the nature of these alien worlds is to measure the ultraviolet (UV), optical or infrared (IR) spectra of their atmospheres.

Measuring the atmosphere of an exoplanet is not an easy task; the atmospheric signals reviewed in this paper are small (typically $10^{-3}$ to $10^{-5}$ as intense as the light from the host star), and the host star usually can not be wholly eliminated from contaminating the measurement.  Moreover, many of the observatories and instruments used to detect and study exoplanetary spectra were not designed for that task.  Consequently, some initial results in this field have proven to be illusory, and have been overturned (e.g., detection of carbon dioxide, Sec. 4.1.2, and carbon planets, Sec. 4.1.4). Nevertheless, thanks to many persistent and talented investigators, we are on the path to a robust and realistic picture of these faint and distant worlds.  Future work using powerful new ground-based and space-borne spectroscopic facilities will expand on that reality, to a degree that will probably exceed our current expectations.

In this review, we discuss the current state of understanding exoplanetary atmospheres, with emphasis on what results we deem to be trustworthy and realistic, what results have been illusory, and what results we think will transition from illusion to reality.  The level of this review is intended to be informative for the astronomer, planetary scientist, or geophysicist who does not specialize in exoplanetary work, but we hope that some of our points will be useful for the exoplanetary specialists.  Because this field moves fast, previous reviews are in danger of becoming obsolete, but Seager and Deming (2010) is still useful in part.   More recently, Burrows (2014) describes some of the highlights in the field, and Marley et al. (2013) review the important topic of clouds and hazes.   Other recent reviews include Crossfield (2015),  Heng and Showman (2015) on atmospheric dynamics, and Madhusudhan et al. (2016).

## 2.0 Prevalence, variety, and stability of exoplanetary atmospheres

We cannot study exoplanetary atmospheres unless we are certain that they exist.  Fortunately, there is clear evidence that atmospheres are very common.  We here discuss their prevalence (Sec. 2.1), their variety (Sec. 2.2) and their stability (Sec. 2.3).

## 2.1 Prevalence of atmospheres on exoplanets

The most reliable inference, with minimal model-dependence, indicating the presence of an atmosphere is when the bulk density is sufficiently low to require a gaseous envelope.  Indeed, that conclusion was evident with the discovery of the first transits of the gas giant HD 209458b (Charbonneau et al. 2000, Henry et al. 2000).  For background, masses of exoplanets are commonly measured using the radial velocity reflex of the host star, or by transit timing variations (TTVs) for those systems where multiple planets transit (pass in front of their star). Transits yield robust radii of exoplanets (Sec. 3.1). The masses and radii for Neptune-sized to super-Earth-sized planets are illustrated in Figure 1, overlaid by lines of constant mass density.

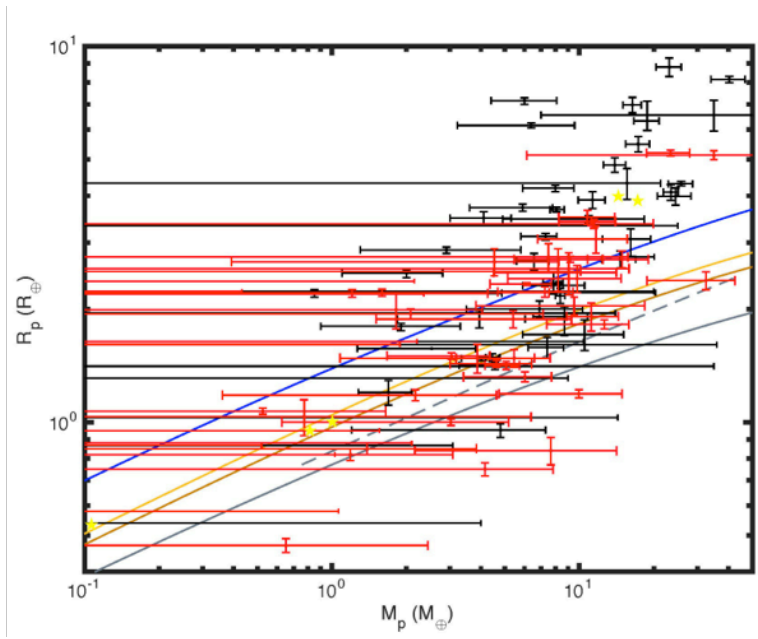

Figure 1. Planet masses and sizes for a range of planets. Shown is the radius (in units of Earth radii, log scale) versus mass (in units of Earth masses, log scale) for super-Earth-sized through Neptune-sized planets, adapted from Rogers (2015). The blue, brown, and gray lines are the loci of planets composed of water ice, silicates, and iron, respectively, from Seager et al. (2007). Since many super-Earths lie above the water ice line (e.g., upper right portion), they must have substantial gaseous envelopes, hence atmospheres. However, planets that lie on or below the locus of water ice planets (e.g., lower left region), do not necessarily host an atmosphere.

Although the conclusion that giant exoplanets have atmospheres is easy, the question becomes more problematic for less massive and smaller exoplanets whose bulk densities are less certain. Fortunately, significant work has been done on this question. Using masses and radii from Kepler plus a minimum of model-dependence, Rogers (2015) concluded that the transition between purely rocky planets (that may or may not have thin atmospheres), and planets having substantial hydrogen envelopes (hence, atmospheres) occurs at about 1.5 $R_\oplus$. Using structural models, Lopez and Fortney (2014) found that radius is a good proxy for the composition of sub-Neptunes, and hence indicative of the nature of the atmosphere. They also conclude that planets smaller than about 1.75 $R_\oplus$ represent a population of rocky super-Earths. Above that size are the mini-Neptunes that have substantial hydrogen-rich envelopes (hence, atmospheres). It is now believed that the transition between planets that must have atmospheres, and smaller rocky planets that might or might not have atmospheres, lies somewhere between 1.5 and 1.7 $R_\oplus$ The nature of atmospheres on rocky planets (super-Earths) is highly uncertain. Many physical processes can produce atmospheric compositions very different from the planet as a whole.

Moreover, we do not securely know a given planet's bulk composition, because there are degeneracies in mass-radius relations that allow planets of different composition and structure to have closely similar masses and radii (e.g., Seager, Kuchner and Hier-Majumder 2007, Swift et al. 2012). Adams, Seager and Elkins-Tanton (2008) pointed out that a variety of compositions can match the average density of Neptune-sized planets, and they found that even a low-mass atmosphere can dramatically increase the radius of the planet (e.g., 0.1-10% of the total mass as atmosphere increases the radius by 5-60% above the gas-free value). Miller-Ricci, Seager and Sasselov (2009) found that super-Earths (rocky or icy planets up to approximately 10 $M_\oplus$) may be able to retain massive hydrogen-rich atmospheres, but many could have high molecular weight atmospheres similar to the terrestrial planets in our Solar System. For a very few rocky planets, we can rule out the existence of significant hydrogen-rich atmospheres based on observations (e.g., TRAPPIST-1b and -1c; de Wit et al. 2016a). Nevertheless, we are making only slow progress on rocky planet atmospheres, due to the lack of observational constraints.

Atmospheres are probably common on small rocky planets, although we currently cannot prove that observationally. The best way to prove the existence of an atmosphere on a small rocky planet is to detect a spectral feature due to a gaseous species. There is currently no such detection that we find to be convincing for planets below the size of Neptune. It will be difficult, but possible, to make such detections in the future, as we describe in Sec. 5. In principle, the existence of atmospheres could also be inferred indirectly. For example, Seager and Deming (2009) point out that the IR phase curve of a close-in tidally-locked planet could indicate the presence of an atmosphere via the longitudinal transfer of heat. However, close-in planets that are volcanic (e.g., via tidal heating) such as claimed for 55 Cnc e (Demory et al. 2016) could potentially have phase curves that mimic the effect of atmospheres.

## 2.2 Variety of exoplanetary atmospheres

We have good evidence that exoplanetary atmospheres vary widely in their properties. This is clearest for transiting giant exoplanets, because their atmospheres are the most easily characterized. Clouds especially have a large impact on transit observations. More than just optically thick clouds, thin hazes of small particles (e.g., produced by photochemistry) can have a large effect on transit observations. Sing et al. (2016) demonstrated that the strength of water vapor absorption measured during transit correlates with a measure of the continuous opacity, in a manner consistent with hazes whose opacity varies by over two orders of magnitude for a sample of 8 giant exoplanets (see Sec. 4.3.1). Clouds also affect the geometric albedo of exoplanets, as measured by the amplitude of secondary eclipse for transiting exoplanets observed at optical wavelengths (where thermal emission does not dominate). While most giant exoplanets have low albedos (e.g., Rowe et al. 2008), due to clear atmospheres that trap and absorb light, Kepler-7b has highly reflective clouds that are distributed non-uniformly, yielding a day side geometric albedo of $0.35 \pm 0.02$ (Demory et al. 2011, see Sec. 4.2.4). We currently do not understand these variations in cloud properties, but they are probably related to irradiation levels, elemental abundances, and surface gravities, among the most important properties.

We are especially interested in the atmospheres of rocky super-Earths. The variety of atmospheres that can exist on them is largely unconstrained by current observations. Transit observations quickly ruled out a moderately clear hydrogen-rich atmosphere for the super-Earth (arguably, mini-Neptune) GJ1214b (Bean et al. 2010). A clear hydrogen-rich atmosphere was also ruled out for the super-Earth HD 97658b (Knutson et al. 2014). While we cannot exclude that super-Earth atmospheres are frequently $H_2$-rich but cloudy, a simpler hypothesis is that super-Earths frequently lose the light element component of their atmospheres. Atmospheric loss is especially plausible for that portion of the population that is very close-in to their host stars, and radial velocity and transit surveys are biased toward finding these close-in planets.

We want to find rocky super-Earths that do have $H_2$-rich atmospheres. They are important for two reasons. First, their atmospheres will have large scale heights, and that makes them easy to characterize by transit spectroscopy if they are at least moderately clear. Second, $H_2$-rich atmospheres are likely to be primordial, and they give us a window into the planet formation process. There appear to be good candidates for such atmospheres. Some Kepler planets have been reported with masses characteristic of super-Earths (e.g., $2M_{\oplus}$ for Kepler-51b, Masuda 2014), and with densities below 0.5 g cm$^{-3}$. The masses for these planets are derived from TTVs, rather than from the radial velocity reflex of the host star—that is too difficult to measure for the relatively faint Kepler stars. Assuming that the TTV masses are correct, these planets would likely have $H_2$-rich atmospheres.

For now, the expected variety of atmospheres on rocky planets comes from theoretical expectations of the stochastic nature of planet formation and volatile delivery in putative protoplanetary disks. Also, secondary atmospheres can result from degassing either during the accretion process, or after the planet is fully formed (Elkins-Tanton and Seager 2008). The composition of degassed atmospheres of super-Earths was modeled by Schaefer and Fegley (2007, 2010) based on chemical equilibrium and chemical kinetics calculations applied to the composition of chrondritic material. These atmospheres can have significant amounts of $H_2$, CO, $CO_2$, and $H_2O$. Schaefer and Fegley (2009) modeled the formation of atmospheres on hot rocky planets that lack volatiles. In that case, the atmospheres are composed of atomic sodium, molecular and atomic oxygen, and SiO.

## 2.3 Stability of exoplanetary atmospheres

Many "hot Jupiters", Neptunes, and super-Earths orbit very close to their host stars; orbital radii of a few times 0.01 AUs are not uncommon. These planets are often losing their atmospheres. Following the discovery of the first exoplanet to orbit a solar-type star (51 Peg b, Mayor and Queloz 1995), Guillot et al. (1996) concluded that such hot Jupiters were stable to classic Jeans escape, and also stable to escape caused by EUV photodissociation of their upper atmospheres. However, atmospheric escape rates are quite sensitive to exospheric temperatures and stellar EUV fluxes, both of which can be very uncertain. Subsequently, Vidal-Madjar et al. (2003) discovered an extended halo of atomic hydrogen surrounding the transiting hot Jupiter HD 209458b, indicating atmospheric mass loss of approximately $10^{10}$ g sec$^{-1}$, and mass loss rates of that magnitude have been inferred for many hot Jupiters (Ehrenreich and Desert 2011). The mass loss is driven by EUV irradiation of the upper atmosphere, that dissociates molecules into fragments with sufficient kinetic energy for atmospheric escape. The resultant perturbation to the hydrostatic equilibrium pressure gradient causes hydrodynamic outflow of the atmosphere.

Although the (in)stability of hot Jupiters to atmospheric mass loss is at least partially understood, the stability of super-Earth atmospheres is still an open question. It's an important question, because we do not actually know whether the hottest and most observable super-Earths have atmospheres, or whether they lose them very easily. Several mechanisms can remove these atmospheres, considered by Heng and Kopparla (2012) for planets below the size of Saturn. Those mechanisms include Jeans escape (Hunten 1990), hydrodynamic outflow powered by EUV radiation (Lammer et al. 2003), and possibly freeze-out (Wordsworth 2015). Freeze-out could occur either in polar regions, or in some instances on the cold side of tidally-locked planets.

In order to further understand the existence and stability of super-Earth atmospheres, it will be necessary to measure their properties observationally. We can suggest two specific techniques to make those measurements. First, thermal emission phase curves (Sec. 3.1.1), as potentially measured using JWST, could reveal the existence of super-Earth atmospheres and constrain their heat capacities. Second, transit spectroscopy of strong resonance lines in the ultraviolet - conducted using HST - is in principle very sensitive to atomic components of the atmosphere. Intrinsically strong UV lines can absorb starlight very efficiently, maximizing the sensitivity for a given atmospheric column density, provided that the star emits a sufficient UV flux.

### 3.0 Contemporary techniques for measuring exoplanetary atmospheres

The study of exoplanetary atmospheres is driven by observations. We here describe the techniques that are used to make the observations, deferring the actual results until Sec. 4.

### 3.1 Combined light

Most measurements of exoplanetary atmospheres to date have been made using the combined light of the planet and host star, specifically without spatially resolving the planet from the star.

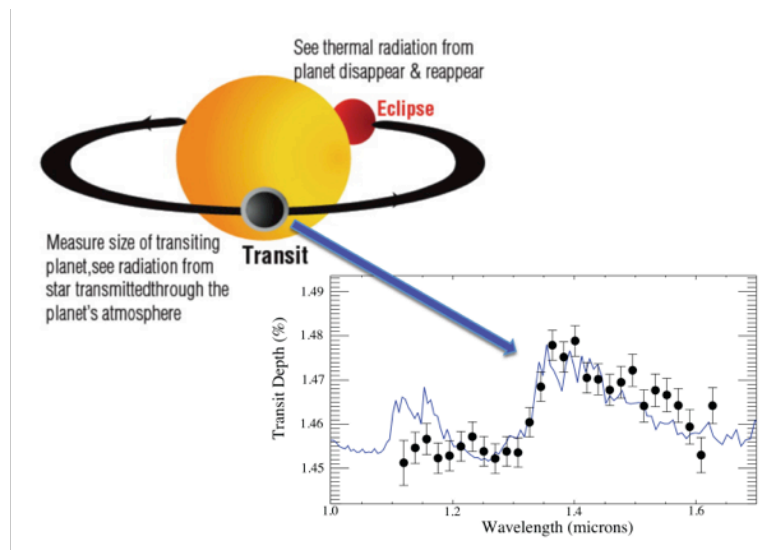

Figure 2. Transit cartoon showing transit and eclipse. An example of an absorption spectrum at transit observed for the exoplanet HD209458b using HST/WFC3 (Deming et al. 2013). As noted in the text, the convention in this field is to plot the absorption spectrum upside-down, because greater absorption increases the planet's transit depth.

### 3.1.1 Transits, eclipses, and phase curves

The vast majority of combined light measurements have been made for transiting planets. The geometry of a transit is shown in Figure 2. Transits occur for a fraction of planets whose orbital plane is aligned (by chance) with our line of sight. Once each orbit, the planet passes in front of the star, which is called the transit. Unless the orbit is highly eccentric, the planet passes behind the star one half of an orbital period later, called the secondary eclipse (or, simply the eclipse, or occultation).

When planets transit their stars, the principal effect is that the disk of the planet blocks light from the star, and the resultant dip in the stellar flux allows us to calculate the radius of the planet. Since the radial velocities for host stars of many transiting planets can be measured to high precision, that tells us the planetary mass. Knowing mass and radius, we then know the bulk density of the planet. As seen by the observer, a tiny fraction of the light from the star passes through the exoplanetary atmosphere during transit. The spectrum of the system then exhibits very weak (typically 200 parts-per-million, ppm) absorption lines or molecular bands due to the planet's atmospheric constituents (Seager and Sasselov 2000). Figure 3 illustrates a modeled and observed spectrum of that type, called a transit spectrum or transmission spectrum. A convention for such spectra is to plot them upside-down, where greater absorption is higher on the figure. The reason for this is that greater absorption makes the planet appear larger because the atmosphere become opaque, and we usually use the apparent radius of the planet as the ordinate on such plots.

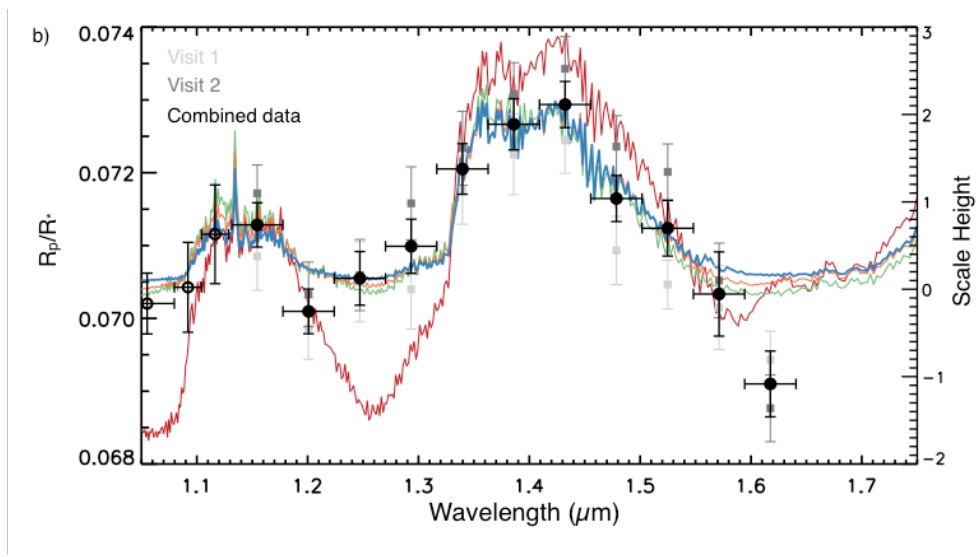

<u>Figure 3.</u> Modeled transmission spectra for a hot Neptune-mass planet from Wakeford et al. (2016b), showing increases in the planet's transit radius (i.e., transit depth) at wavelengths of prominent molecular absorption. Different modeled curves have different cloud properties and different carbon-to-oxygen ratios. The observations (black and gray points) are HST observations, discussed in Sec. 4.2 Note the strong absorption peak due to water vapor at 1.4 microns, and note the consistency between two independent sets of observations (visits).

When the planet passes behind the star at secondary eclipse, the light reflected or emitted by the planet is blocked by the star. The amplitude of the resultant dip in total light measures the light due to the planet (Charbonneau et al. 2005, Deming et al. 2005). The amplitude of secondary eclipses for most planets is greatest in the IR, and an example of an IR secondary eclipse is shown in Figure 4. The secondary eclipse can be measured using either photometry in broad bands, or using spectroscopy. Photometric eclipses are relatively easy to measure, but hard to interpret (as we discuss in Sec. 4.2), whereas spectroscopic eclipses are easier to interpret, but harder to measure because the number of photons available is often too scant to obtain sufficient signal-to-noise for spectroscopic eclipses.

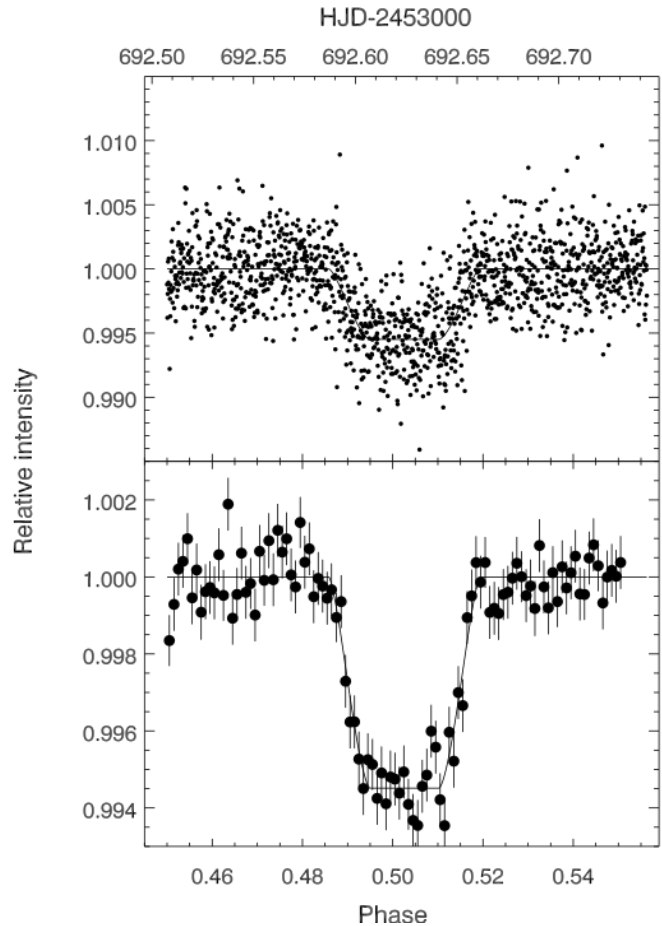

Figure 4. Secondary eclipse of the hot Jupiter HD189733b at 16 microns wavelength, from Deming et al. (2006). The top panel shows the original unbinned data, the lower panel bins the data to illustrate the eclipse more clearly.

### 3.1.2 Eclipse mapping

When a planet passes behind it's star, we can in principle exploit that event to learn about the distribution of light across the planetary disk.  That is possible because the scale height in the *stellar* atmosphere is typically much less that the radius of the planet, especially for giant planets like HD 189733b (whose eclipse is shown in Figure 4).  So the star is effectively a "knife edge" compared to the planetary disk. The zero-th order effect during eclipse ingress and egress is the shape of the planet (Seager and Hui 2002).  However, with sufficient signal-to-noise during the ingress and egress of the planet, the measurements could be inverted to produce a map of the intensity emitted by the planet as a function of the position on the planetary disk (Majeau et al. 2012, de Wit et al. 2012).

Close-in planets that have tidally-locked rotation will present the same hemisphere toward the star as they orbit.  In that case, the angular phase of the star-facing hemisphere goes through a complete revolution relative to our line of sight once per orbit.  When the (hot) star-facing hemisphere is facing us, the IR signal from the planet is near maximum, and the signal is near minimum when the star-facing hemisphere is turned away from us.  That varying signal is called the phase curve, and the phase curve can be combined with measurements during ingress and egress to improve the intensity map of the planet (Cowan and Agol 2008).   The first phase curve measurement (Knutson et al. 2007) showed that the hottest point on the disk of HD189733b was displaced from the sub-stellar point in a direction and magnitude that is consistent with advection of heat by zonal winds (Showman et al. 2008).

### 3.1.3 Doppler deconvolution

Planetary atmospheres can be measured in combined light, even for planets that do not transit.  Radial velocity measurements of a host star can detect the existence of a planet, and can measure the orbital period.  In that case, the combined light spectrum of the system contains the spectrum of the planet, that is Doppler-shifted with a known period and phase, but an unknown amplitude.  By convolving a theoretical template spectrum of the planet with high signal-to-noise spectra of the combined light of the system, the amplitude and phase of the planet's spectral variation can be identified via peaks in the cross-correlation function.  We call this technique Doppler deconvolution.  Pioneering Doppler deconvolution attempts to detect hot Jupiter atmospheres were first reported by Charbonneau et al. (1999), Collier-Cameron et al. (1999), and Wiedemann et al. (2001).   Collier-Cameron et al. claimed probable detection of a signal from the hot Jupiter tau Bootis b, a result that proved to be illusory.  However, the advent of the CRIRES instrument to obtain stellar spectra at very high resolving power has brought the Doppler deconvolution results into the realm of reality, as we discuss in Sec. 4.2.

### 3.2 Planet-star spatially resolved spectroscopy

The most intuitive way to obtain the spectrum of an exoplanet's atmosphere is to spatially resolve the planet from the star, and direct the light from the planet into a spectrometer.  This is technically difficult to accomplish, but has been increasingly successful in recent years.  The planetary system most amenable to this technique are those that have giant planets orbiting at large distances from their host star.  At large orbital distances, planets will not reflect significant amounts of star light, nor will they be sufficiently heated by stellar irradiation to be detected in the IR. However, young giant planets can produce IR radiation from their heat of formation. Figure 5 shows a spatially resolved system of young giant planets orbiting at large distances from the hot star HR8799 (Marois et al. 2008, Currie et al. 2014). Spectra of these planets, and of a similar planet orbiting Beta Pictoris (Chilcote et al. 2015) have been obtained both from the ground and from HST, as we discuss in Sec. 4.2.2.

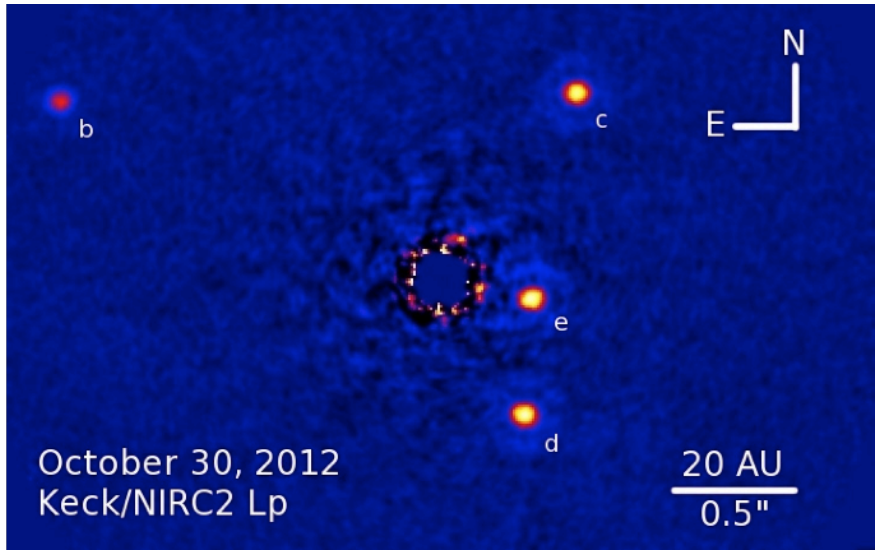

Figure 5. Image of the exoplanets orbiting the star HR8799 (Marois et al. 2008), using the NIRC2 instrument on the Keck telescope (Currie et al. 2014).

A variation on spatially resolved spectroscopy has been proposed and simulated (e.g., Snellen et al. 2015) wherein coronography is used to improve the planet-to-star contrast ratio without fully separating star and planet. Spectra with the star partially rejected in this manner are then analyzed using the Doppler deconvolution technique to extract the exoplanetary spectrum. This hybrid method may be especially advantageous for non-transiting planets in nearby systems like Proxima Centauri only 1.3 parsecs distant (Anglada-Escude et al. 2016, Lovis et al. 2016). The nearer the system to Earth, the greater the planet-star separation angle and the more photons we collect. So both the effectiveness of the coronography, as well as the signal-to-noise for Doppler deconvolution (via the photon flux), are maximized for nearby systems. Hence this hybrid method is doubly weighted to nearby exoplanetary systems. However, we caution that the method has yet to be proven on real data, as it requires a planet with a large Doppler shift and a planet-star spatial separation that is at least minimally resolvable.

### 3.3 Factors affecting the measurements

Both combined light and spatially resolved techniques are subject to interference by clouds in the atmosphere of the planet, and by the host star, as we now discuss.

### 3.3.1 Aerosols

Transmission spectroscopy of a transiting planet relies on light from the star being transmitted through the planetary atmosphere, except at wavelengths where atomic and molecular absorptions block the transmission. When aerosols (clouds and haze) occur, there are two implications. First, the absorption spectrum of the planet can be completely blocked (e.g., Kreidberg et al. 2014a), or the absorption spectrum can be weakened. The latter case introduces a degeneracy between the abundance of the absorbing species and the distribution and optical thickness of the cloud layers (Benneke 2016, Line and Parmentier 2016). The effect is more severe for transit than for eclipse spectroscopy, because of the long slant paths traversed by stellar light at transit. The amplitude of spectral features in transmission spectroscopy is sensitive to the form of the cloud, i.e. whether clouds form an optically thick layer occurring at specific pressure levels, or whether they form a uniformly distributed haze of small particles. The first transit spectroscopy of atomic sodium in the giant planet HD 209458b (Charbonneau et al. 2002) was found to be weaker than predicted by clear atmosphere models (Seager and Sasselov 2000). The transit spectrum of water vapor in the same planet was similarly found to be weaker than predicted from clear solar abundance models (Deming et al. 2013). One interpretation of that weakness is a low abundance of oxygen (Madhusudhan et al. 2014). Given the effect of clouds, low oxygen abundances can be illusory. Spectroscopy of HD209458b at eclipse—where clouds have less impact—is consistent with solar oxygen abundance for that planet (Line et al. 2016). We further discuss abundances in exoplanet atmospheres in Sec. 4.3.1. For directly imaged planets, clouds affect the measurements differently. Since directly imaged planets are spatially resolved from their star, they can be observed at many different orbital phases. This translates to viewing the planetary atmosphere at different angles relative to the sub-stellar point, and we expect varying cloud properties as a function of viewing angle due to weather patterns. Directly imaged planets are closely related to brown dwarfs, where there is abundant evidence for patchy clouds (e.g., Zhou et al. 2016, Crossfield et al. 2014). In principle, the variation of cloud properties with viewing angle and rotation is an advantage, not a hindrance to breaking degeneracies between clouds and abundances.

### 3.3.2 The host star

The zeroth order effect of the host star is that its large photon flux adds noise to spectra of the planetary atmosphere. Combined light techniques deal with that problem by seeking very high signal-to-noise and dynamic range in the measurements, that can be acquired with conventional instrumentation. Direct imaging uses optical technology to block the light from the host star, thus minimizing the impact of its photon noise, but such measurements are technologically challenging. Moreover, beyond the photon noise from the host star, other stellar effects must be considered. The most prominent of these is stellar magnetic activity, that can cause temporal variability and spatial inhomogeneity on the stellar disk. Combined light measurements are much more strongly affected than direct imaging.

Temporal variability of the host star can in principle affect transit and eclipse measurements, but in practice the time scale of stellar variability is dominated by stellar rotation, and tends to be of longer duration (typically, days) than a transit or eclipse (typically, hours), thus minimizing its effect on those measurements. Phase curve measurements are more sensitive to stellar variability than are transits and eclipses, but monitoring of the star at optical wavelengths provides leverage to correct for stellar variations (Knutson et al. 2012).

The spatial inhomogeneity of the stellar disk is problematic for transit measurements (but not for eclipses). For example, as a planet transits a star it can occult star spots. Large single spots can be obvious in the transit light curve (e.g., Figure 2 of Deming et al. 2011), but small star spots can have a significant aggregate effect while remaining under the threshold for detection individually. Moreover, star spots that are *not* transited can have the greatest impact. Spots often occur at "active latitudes" and can be missed by a planet that transits on the stellar equator. In that case, the star will appear redder during transit than out of transit, because the cool unocculted star spots will have greater relative weight during transit. Increased redness during transit can mimic the effect of scattering by small particles in the exoplanet's atmosphere, as demonstrated by McCullough et al. (2014). Fortunately, the effect of star spots can be discriminated from planetary atmospheric scattering by evaluating the level of stellar activity using proxy indicators, and by the strengths of the alkali absorption lines in the planet's transmission spectrum (Nikolov et al. 2014).

### 3.4 Techniques not yet fully exploited

We point out the potential for exoplanet atmospheric characterization using two techniques that have not been fully exploited to date: polarization and refracted light.

### 3.4.1 Polarization

Exoplanetary atmospheres are predicted to scatter star light with an efficiency that depends on the polarization state (Seager, Whitney and Sasselov 2000). The magnitude of the polarization signal is predicted as typically tens of ppm, but is very sensitive to the nature of the exoplanetary atmosphere. Although tens of ppm is a very small signal, polarization measurements are sensitive because the starlight itself is unpolarized, making the planet-star contrast favorable for combined-light techniques, and the observations are typically conducted by rapid switching between opposite polarizations, which can often eliminate instrument errors (but not intrinsic photon noise). Polarization measurements are a good example of how exoplanet measurements often transition from illusion to reality (see Sec. 4.1.1). These measurements could in principle benefit greatly from the advent of the Extremely Large Telescopes (ELTs) at ground-based observatories (Sec. 5.1), because large fluxes are needed to drive the noise levels down sufficiently.

### 3.4.2 Refracted light

During a transit, star light is refracted as it passes through the atmosphere of the exoplanet (Hui and Seager 2002). Refraction is a familiar phenomenon for solar system planet occultations of background stars, and Figure 6 shows the symmetry between refraction in the solar system and in exoplanets.

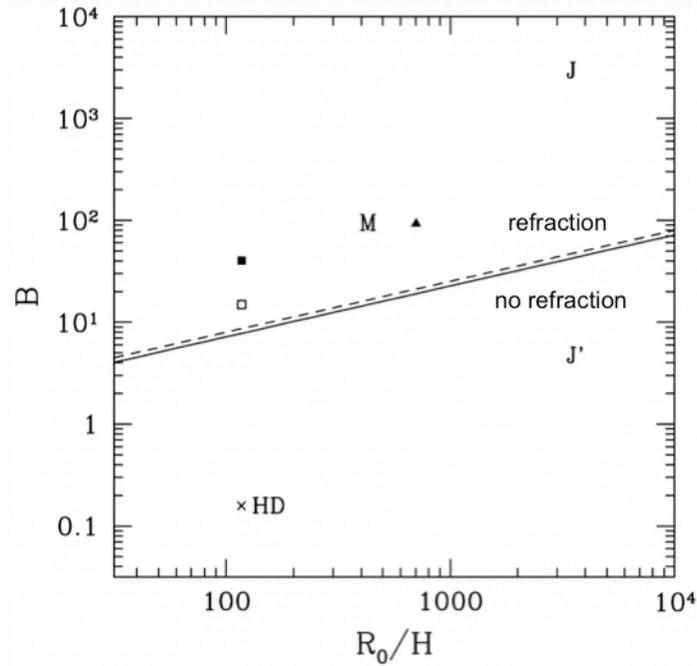

Figure 6. Overview of exoplanet atmosphere refraction. The separation of the strong atmospheric refraction regime (likely existence of refraction) from the weak atmospheric refraction regime for a combination of planet parameters is delineated by the solid and dashed lines (where the dashed line is for an elliptical planet). $B$ (y-axis) is a refraction deflection angle scaled by ratios of distances, proportional to $(\rho_0/H)D_{LS}D_{OL}/(D_{OL} + D_{LS})$, where $\rho_0$ is the density at which the atmosphere begins to be transparent (i.e. low optical depth). We borrow from gravitational lensing terminology (since refraction is a "lensing" phenomenon); $D_{LS}$ is the distance between the planet and star, and $D_{OL}$ is the distance between the observer and the planet. O refers to observer, S source (the star) and L lens (the planet atmosphere). The distances show the symmetry between atmosphere refraction of distant starlight by solar system planets, and exoplanet atmospheric refraction of their host source stars. $R_0/H$ (x-axis) is the ratio of planet radius to atmospheric scale height. Symbols illustrate likelihood of refraction for different cases: HD 209458b (HD), Mars (M), Jupiter (J). The symbol J' uses the parameters for Jupiter if it were observed at a wavelength with 600 times stronger absorption, than at visible wavelengths, showing there is no refraction when absorption dominates. Figure from Hui and Seager (2002).

In general, refraction will occur at a wavelength range where absorption by molecules and absorption and scattering by hazes is absent, and where combined effects of the planet-star separation, atmospheric scale height, and the wavelength-dependent refraction are favorable (Hui and Seager 2002; see Figure 6 for a concise formula describing the combinations).

Refraction can limit the range of atmospheric pressures that are sensed via transit spectroscopy (Garcia Munoz et al. 2012), and can affect the atmospheric "continuum", so it must be carefully accounted for when interpreting transit spectra. The magnitude of refracted light depends on the scale height of the atmosphere (Figure 6; Hui and Seager, 2002; Sidis and Sari 2010). More important than the limiting effects of refraction are its potential as a diagnostic of conditions in exoplanetary atmospheres (Misra and Meadows 2014). Future transmission spectroscopy of exoplanetary atmospheres will benefit by finding the clearest atmospheres with the largest scale heights, because those planets will have the most easily measured transmission spectra. Detection of refracted light can indicate such planets. Light that is refracted deep in the exoplanetary atmosphere will tend to reach us before the transit begins, and after it ends, producing "shoulders" on the transit light curve (Hui and Seager 2002). Refraction shoulders have not been unequivocally detected to date, but would indicate a planet with a clear and extended atmosphere - an ideal candidate for transmission spectroscopy, e.g. using JWST (Sec. 5.2).

## 4.0 Status of results for exoplanetary atmospheres

We now summarize the principal results on characterization of exoplanetary atmospheres, beginning with results that we regard as illusions (Sec. 4.1), followed by results that represent reality (Sec. 4.2), and then we highlight two particular themes that are still emerging (Sec. 4.3). Figure 7 summarizes our opinion regarding the illusion vs. reality status of selected topics, in the form of a "scorecard".

| Science | Illusion? | Goal? | Transition | Reality? | Comment | Sec. |
|---|---|---|---|---|---|---|
| Polarization | X | | ⇒ | | Small amplitude = very difficult | 4.1.1 |
| $CH_4$ and $CO_2$ in transiting planets | X | | ⇒⇒⇒ | | JWST spectroscopy | 4.1.2 |
| Temperature inversions | X | | ⇒ | | Maybe too optically thin | 4.1.3 |
| Carbon planets | X | | ⇒⇒ | | JWST spectroscopy | 4.1.4 |
| Atomic absorption in transiting giant planets | | | | X | HST and ground-based | 4.2.1 |
| Water vapor in transiting giant planets | | | | X | Many HST/WFC3 measurements | 4.2.1 |
| Water vapor in transiting super-Earths | | X | ⇒⇒ | | Maybe HST, but also JWST | 4.2.1 |
| Water and CO in close-in giant planets | | | | X | Doppler deconvolution; ground-based | 4.2.1 |
| Water & CO in directly imaged planets | | | | X | Ground-based adaptive optics + HST | 4.2.2 |
| Efficiency of heat redistribution for tidally locked planets | | | | X | Spitzer phase curves + JWST? | 4.2.3 |
| Optical geometric albedos of transiting giant planets | | | | X | MOST + Kepler | 4.2.4 |
| Bond albedos of transiting giant planets | | | | X | Spitzer + JWST | 4.2.4 |
| Inhomogenous clouds on transiting giant planets | | X | ⇒ | | JWST + TESS? | 4.2.3 |
| Albedos of transiting super-Earths | | X | ⇒ | | Photon-starved | 4.3.2 |
| Abundances | | X | ⇒⇒ | | JWST eclipse spectra | 4.3.1 |

Figure 7. "Scorecard" summary of illusions, realities, and goals for exoplanetary atmospheric science. The "Transition" column gives our subjective estimate of how rapidly each topic will transition from illusion or goal, to reality: the more arrows the faster the result will become reality.

## 4.1 Illusions

Exoplanet investigators are often dealing with signals at the limits of detectability. It is natural, even inevitable, that many of the earliest results will be illusory. Our discussion is not meant as critique of the early work, but to illustrate how the field moves from the first tentative and often incorrect conclusions to a firm and realistic view of exoplanetary atmospheres.

### 4.1.1 Polarization

The first report of exoplanet polarization (Berdyugina et al. 2008) indicated a polarization signal from HD189733b amounting to about 200 ppm, much larger than expected. Subsequent measurements by Wiktorowicz (2009), and Bott et al. (2016) convince us that the initial result was an illusion, whereas much lower levels of polarization, consistent with theoretical predictions, are the reality. Much more observational work needs to be done on this topic.

### 4.1.2 Molecules in transmission

The first detection of an exoplanet atmosphere (Charbonneau et al. 2002) was *not* an illusion. At the time, it was greeted with private skepticism, as important new results frequently are. Nevertheless, it has been confirmed by many subsequent measurements. That early success stimulated additional work on transmission spectroscopy, focused in the near-IR where molecular absorptions are potentially detectable. Spectroscopy using the NICMOS instrument on HST indicated detection of water vapor and methane (Swain et al. 2008), as well as carbon dioxide (Tinetti et al. 2010). Unfortunately, the NICMOS instrument has particularly severe wavelength-dependent instrumental signals that interfere with transit spectroscopy. Swain et al. (2008) recognized those instrumental signatures, and attempted to remove them by decorrelating the transit signals versus parameters of the instrument and detector (e.g., position and orientation of the spectrum). Subsequently, Gibson et al. (2011) questioned the reality of the molecular detections using NICMOS, because they found that the absorption signal depended on how the decorrelations were performed, i.e. on the instrument model they used. Since there was no *physical* reason to prefer one instrument model over another, Gibson et al. concluded that there was no firm evidence for the reality of the molecular detections, and large errors for NICMOS spectroscopy were also inferred by Crouzet et al. (2012). Moreover, for the exoplanet XO-1b, the NICMOS transmission spectrum derived by Tinetti et al. (2010) was contradicted by results using WFC3 spatial scan spectroscopy (Deming et al. 2013), the latter of which is now widely accepted as reliable. However, Waldmann et al. (2013) concluded that the NICMOS results were themselves reliable, based on a blind (hence, unbiased) spectrum extraction technique.

Since the NICMOS instrument is no longer operational, and WFC3 on HST is producing robust transmission and eclipse spectroscopy (Sec. 4.2), and JWST is imminent (Sec. 5.2), the reliability of the NICMOS detections is somewhat moot. Our opinion is that most of the NICMOS results (detection of methane and carbon dioxide) are illusory. However, based on Figure 1 of Gibson et al. (2011), consistent absorption peaks near 1.8 $\mu$m suggest that water vapor may indeed have been detected, albeit with large errors. In Sec. 4.2, we discuss the reality of water vapor in giant exoplanets, as observed using the WFC3 instrument on HST, and other facilities.

### 4.1.3 An archetype temperature inversion

Similar to the Earth's stratosphere, most planets in the Solar System exhibit temperature inversions at some (usually high) level in their atmospheres. The upper atmosphere of Jupiter is heated to temperatures in excess of 1000 K (Atreya and Donahue, 1979). The mechanism of Jupiter's heating is not understood, but solar particle precipitation, dissipation of atmospheric gravity waves, and other mechanisms have been investigated as possible drivers. For close-in exoplanets, strong stellar irradiation provides a viable energy source to drive temperature inversions, and searching for possible inversions has been an active topic in exoplanetary science.

Secondary eclipse photometry using the Spitzer Space Telescope yields the flux emitted by exoplanets at wavelengths between 3.6 and 24 micrometers. One of the earliest secondary eclipse results was photometry of the hot giant planet HD209458b. Those results showed a greater eclipse depth at 4.5 and 5.8 micrometers compared to the prediction of a standard model (Knutson et al. 2008). The inference was that the temperature profile in the deep exoplanetary atmosphere was inverted, with temperature rising with height instead of falling as in the model. The rising temperature would produce spectral emission features in the 4.5 and 5.8 micrometer bandpasses, accounting for their excess eclipse depths. Weaker evidence was found for temperature inversions in other strongly-irradiated exoplanets (e.g., XO-2b, Machalek et al. 2009, and WASP-18b, Nymeyer et al. 2011). However, Diamond-Lowe et al. (2014) re-analyzed the original photometry, also including newer and better photometry for HD209458b, and concluded that there was no evidence for a temperature inversion, and that has been recently confirmed using HST spectroscopy (Line et al. 2016). The illusion in this case occurred because the original Spitzer photometry was obtained and analyzed using early techniques that are not as reliable as current methods. Spitzer photometry, and methods to analyze it, have steadily improved with time. Recently, Ingalls et al. (2016) compared multiple parallel analyses of both real and synthetic Spitzer photometry, and demonstrated consistency and reliability in the best current techniques.

We now realize that the archetype evidence for a temperature inversion (in HD209458b) was an illusion. Yet, planetary atmospheric temperature inversions are a natural consequence of strong stellar irradiation, and there is other evidence for their existence (e.g., Haynes et al. 2015). Hence, the reality of temperature inversions as a general phenomenon is still an open question.

### 4.1.4  Carbon planets

Oxygen is cosmically more abundant than carbon, and most stars have a carbon-to-oxygen (C/O) ratio less than unity (Fortney 2012, Brewer et al. 2016). The C/O ratio in planets is important because it has a major effect on their atmospheric chemistry. There is precedent for believing that planets can be enriched in carbon. Lodders (2004) argued for the formation of Jupiter in a carbonaceous, rather than icy, environment. Kuchner and Seager (2005) suggested that some super-Earth exoplanets could form primarily from silicon carbide and other carbon compounds, rather than from silicon dioxide, as long as the local planet-forming environment has C/O > 1. Models of protoplanetary disks predict that the C/O ratio should vary between planets at different distances from the host star, because water vapor and carbon monoxide have different condensation temperatures (Oberg et al. 2011). Hence, carbon planets are a plausible concept.

Secondary eclipse photometry using the Spitzer Space Telescope yields the flux emitted by the planet over broad photometric bands. In principle, the nature of the planet's molecular absorption spectrum can be inferred from eclipse depths in a few photometric bands. On this basis, Madhusudhan et al. (2011) concluded that the hot transiting planet WASP-12b exhibited a C/O ratio in its atmosphere exceeding unity, i.e. it was a carbon planet. Nevertheless, a high C/O ratio for WASP-12b has not been confirmed by subsequent work (Kreidberg et al. 2015), and limits have been placed on C/O in other planets (Line et al. 2014, Benneke 2016). Madhusudhan and Seager (2009) demonstrated that photometry in Spitzer bands would have been sufficient to constrain basic properties of thermal structure and molecular abundances in hot Jupiters, provided that the error bars were appropriately assigned. However, Hansen et al. (2014) claimed that detection of molecular absorptions using Spitzer photometry are illusions in almost all planets, because the level of systematic error is larger than previous investigators estimated.

The Spitzer analyses have improved with time, as we have gained more insight into the properties of the data. Given the quality of recent Spitzer photometry demonstrated by Ingalls et al. (2016), a re-analysis of Spitzer eclipses should yield a realistic overview of molecular absorption in hot giant exoplanets, as we note in Sec. 4.2.1.

## 4.2  The current reality in sensing exoplanetary atmospheres

Although we have mentioned several illusions that developed in the study of exoplanetary atmospheres, the realities have now exceeded the illusions by far. Highlights of these realities include measurement of atomic and molecular species in the transmission spectra of giant planets, pushing toward super-Earths (Sec. 4.2.1). The technique of high contrast direct imaging is advancing rapidly (Sec. 4.2.2), obtaining the first estimates of abundances in young giant planets at distances beyond the water frost line. Phase curves have been observed for multiple giant planets, yielding fundamental insights into their atmospheric dynamics (Sec. 4.2.3). Photometry at secondary eclipse has measured albedos for close-in giant planets to super-Earths, with implications for scattering and cloud formation in their atmospheres (Sec. 4.2.3).

### 4.2.1 Spectra of giant planets to super-Earths

An exciting reality for our understanding of giant exoplanets is that we can now measure their transmission spectra robustly, albeit for very narrow wavelength regions. The temporal order of the measurements has proceeded—not entirely by coincidence—from the simple (atomic lines) to the more complex (molecular bands).

Absorption during transit in strong atomic lines was an early prediction of models (Seager and Sasselov 2000) and was the first solid observational detection of exoplanetary atmospheres (Charbonneau et al. 2002). Both sodium and potassium lines have been measured using space and ground-based observatories (Redfield et al. 2008, Jensen et al. 2011, Sing et al. 2012, 2015, Nikolov et al. 2014, Wilson et al. 2015). Because the absorptions are due to intrinsically strong resonance lines, they probe low pressure levels and relatively high altitudes in the exoplanetary atmosphere. Thus, we learn about the height distribution of clouds (Heng 2016), and potentially about the atmospheric temperature profile up to the thermosphere (Huitson et al. 2012, Vidal-Madjar et al. 2011).

Not only atomic lines, but also water vapor are now detected robustly in both transit and eclipse, using HST/WFC3 grism spectroscopy. Water vapor is routinely measured in giant planets during transit (Deming et al. 2013, Huitson et al. 2013, Mandell et al. 2013, McCullough et al. 2014, Wakeford et al. 2013, 2016a,b, Kreidberg et al. 2014b, 2015, Evans et al. 2016), and also at secondary eclipse (Crouzet et al. 2014, Kreidberg et al. 2014b, Line et al. 2016). An example of water vapor detected during transit is shown in Figure 3. Planets as small as Neptune (Fraine et al. 2014) have yielded water absorption, and ongoing HST programs are pushing toward detection of water vapor and other molecules in super-Earths. We discuss one emerging scientific theme of these observations in Sec. 4.3.1.

Not only HST, but also Spitzer and ground-based observatories have detected molecular absorptions in hot giant planets. A Spitzer water absorption spectrum for the archetype transiting planet HD189733b by Grillmair et al. (2008) was re-visited and confirmed by Todorov et al. (2014), including new data not analyzed by Grillmair et al.

The Doppler deconvolution technique (Sec. 3.1.3) has achieved clear detections of water vapor from the ground (Birkby et al. 2013, Lockwood et al. 2014), and also carbon monoxide (Snellen et al. 2010, Brogi et al. 2012, 2014) in both transiting and non-transiting hot giant planets. This cross-correlation technique yields the orbital velocity, hence mass, of non-transiting planets, and with sufficient signal-to-noise can determine the magnitude of high altitude winds in the exoplanetary atmosphere (Snellen et al. 2010). Also, the amplitude of the cross-correlation peak can in principle yield molecular abundances (Birkby et al. 2013).

An unresolved issue is the degree to which secondary eclipse photometry in broad bands (e.g., using Spitzer) can reveal the presence of molecular absorptions. The point is important because Spitzer bands, unlike HST/WFC3, are very sensitive to carbon-containing molecules like methane and carbon monoxide. Spitzer photometry is already needed for day-side temperature measurements. Given the recent Ingalls et al. (2016) demonstration of reproducibility in recent Spitzer photometry, construction of better infrared color-magnitude diagrams (e.g., Triaud 2014) has the potential to constrain molecular absorptions in hot Jupiter atmospheres.

### 4.2.2 Spectra of directly imaged, young giant planets

The most obvious method to obtain spectra of exoplanetary atmospheres is to spatially resolve a planet from its host star, and direct its radiation into the entrance aperture of a spectrometer. However, spectroscopy via direct imaging is not technologically feasible for most planets, due to their extreme proximity to their stars. Fortunately, advances in ground-based adaptive optics (e.g., Macintosh et al. 2007), applied to young systems of planets that are self-luminous at great distances from their stars, makes this technique possible in special cases. The discovery of a system of planets orbiting the young star HR8799 (Marois et al. 2008) has spawned multiple spectroscopic investigations.

Compared to transiting planets, spectroscopy by direct imaging should be less sensitive to the stability of the spectrometer, since these observations do not rely on analyses of time series data. Instead, these planets tend to be very faint, and the observations are sensitive to the stability of the background subtraction. Large telescopes are often required to collect sufficient photon flux for useful spectra. Successful spectroscopy of the young planets orbiting HR 8799 has been obtained by several teams (Konopacky et al. 2013, Ingraham et al. 2014, Rajan et al. 2015). Konopacky et al. measured absorption due to of water vapor and carbon monoxide in HR8799c (see Figure 8), and found evidence that the gas accretion that formed this planet was characterized by a greater than stellar C/O ratio. A greater than solar C/O ratio is consistent with the formation of HR8799c at a distance exceeding the water frost line, but within the frost lines for carbon monoxide and carbon dioxide. Ingraham et al. (2014) used the Gemini Planet Imager to obtain spectra of both HR8799c and d, and found evidence for thick, patchy clouds. Rajan et al. (2015) obtained near-IR photometry (λ < 1.5 microns) from HST/WFC3 to measure the spectral energy distribution of HR8799b and c, demonstrating good agreement with atmospheric models. Todorov et al. (2016) derived water vapor abundance in the directly imaged planet κ And b, and related that to water abundances in other giant planets, both from direct imaging and transiting planets. Comparative studies of that type have the potential to reveal abundance signatures of planet formation and migration, but with the caveat that atmospheric abundances may differ widely from bulk abundances (see Sec. 4.3.1).

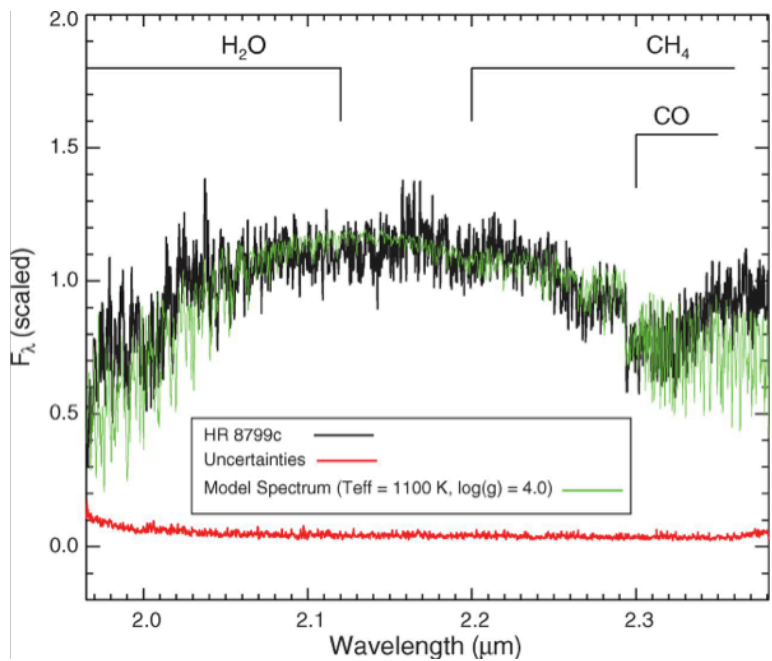

Figure 8. Spectrum of a directly imaged planet, HR8799c, orbiting less than 40 AU from the early-type star HR8799, from Konopacky et al. (2013). The observed spectrum is the black line; the green line is a model. Note the bandhead of carbon monoxide near 2.3 microns, and the absorption of water vapor sloping downward at the shortest wavelengths (on the left). The region of possible methane absorption is marked, but Konopacky et al. did not claim detection of methane features in this spectrum.

### 4.2.3 Phase curves, aerosols, and longitudinal distribution of heat

Spitzer secondary eclipse amplitudes constrain the day side temperature, which can indicate the degree of heat re-distribution even in the absence of a full phase curve measurement (Cowan and Agol 2011). Spitzer secondary eclipse depths are sensitive to the nature of the temporal baseline due to changing detector sensitivity, because eclipses are usually observed for only an hour or two before and after eclipse. In contrast, phase curve measurements cover a large fraction of the planet's orbit, and the phase curve amplitude is minimally sensitive to instrumental baselines, as long as the observations are continuous, or nearly continuous.

Investigators using primarily Spitzer, but also Hubble, have mapped the longitudinal distribution of temperature on tidally-locked hot giant planets. Spitzer phase curves have been measured for eight giant exoplanets, some at multiple wavelengths (Cowan et al. 2012, Crossfield et al. 2012, Knutson et al. 2007, 2009, 2012, Laughlin et al. 2009, Lewis et al. 2013, 2014, Maxted et al. 2013, Zellem et al. 2014, Wong et al. 2015, 2016, de Wit et al. 2016b). Measuring phase curves using spectroscopy (as opposed to photometry) is difficult because of lower signal-to-noise when the spectrum is dispersed. Nevertheless, for the hottest planets such as WASP-43b, HST can measure spectroscopic phase curves (Kreidberg et al. 2014b, Stevenson et al. 2014). An example of a phase curve measured using HST is shown in Figure 9.

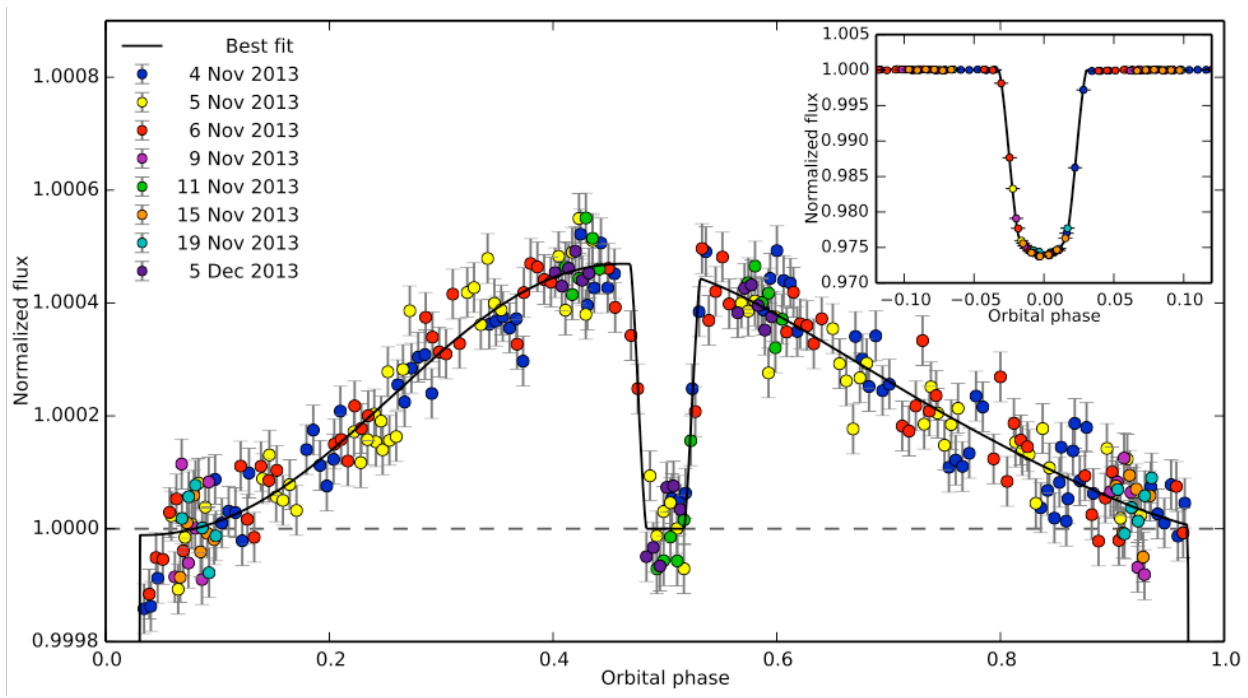

Figure 9. Hubble phase curve of the hot Jupiter WASP-43b from Stevenson et al. (2014). The secondary eclipse is shown in the center at orbital phase 0.5. The increase in flux near the eclipse occurs because the planet is turning its strongly heated day side toward our direction. The inset shows the transit at orbital phase 0.0.

Phase curve observations, especially spectroscopic ones, are potentially rich in information on the dynamics of the exoplanetary atmosphere. The most reliable aspect of a phase curve is its amplitude, that constrains the fraction of stellar irradiation that is absorbed (i.e., the Bond albedo), and the degree of heat re-distribution. Currently, Bond albedos for hot Jupiters inferred from phase curves are higher (by about 0.25) than are geometric albedos from optical measurements. That difference may indicate higher cloud reflectivity in the near infrared (Schwartz and Cowan 2015). More subtle aspects, such as the phase shift in the peak of the curve, are more difficult to measure but the results to date have been consistent. Comparisons between observed phase curve shifts and radiative hydrodynamic models can yield information on the zonal wind speeds, via the relative magnitude of the advective and radiative time scales (Showman et al. 2008). Not only the longitudinal advective time scale, but also the vertical advective time scale are important in this comparison (Perez-Becker and Showman 2013). For planets whose rotation is tidally-locked in a circular orbit, it is not possible to independently measure their radiative and dynamical time scales. However, planets in eccentric orbits like HAT-P-2 (Lewis et al. 2013, 2014; de Wit et al. 2016b) can be used to break this degeneracy. Phase curve observations and models can also inform us of the atmospheric conditions that enable the formation of clouds (Kataria et al. 2016, Parmentier et al. 2016). Moreover, optical wavelength phase curves can sense clouds directly, and infer properties of their spatial distribution, via their reflection of starlight. Those observations indicate that clouds on hot giant planets can be highly inhomogenous in longitude (Demory et al. 2013, Esteves et al. 2013), but the cause of the inhomogenity is not yet understood.

### 4.2.4 Albedos of transiting giant planets

The albedo of a planet informs us concerning the nature of its atmosphere, especially concerning cloud formation. Clouds can often increase the albedo, whereas molecular scattering in a clear atmosphere can make a planet very dark. Optical wavelength geometric albedos of close-in transiting planets can be determined from their secondary eclipses, after correction for thermal emission. Because the signals are small (<< 100 ppm), space-borne observations are required to attain the requisite precision. Precise photometry from the MOST mission enabled Rowe et al. (2008) to establish a 3-sigma upper limit of 0.17 for the geometric albedo of the hot giant planet HD209458b. Kepler photometry established that low albedos are common for hot giant planets (Kipping and Spiegel 2011, Gandolfi et al. 2013, Morris et al. 2013), but some are bright due to highly reflective clouds (Demory et al. 2011). For hot giant planets, the observational picture of primarily low albedos, punctuated by occasional bright clouds, seems to represent reality. Models (Marley et al. 1999, Sudarsky, Burrows and Pinto, 2000) predict a wide variety of albedos for hot giant planets, depending on their equilibrium temperatures. The hottest and clearest atmospheres are predicted to have low albedos, so the darkness of a cloud-free atmosphere is understood.

### 4.3 Current unresolved questions

The nature of exoplanetary atmospheres is sufficiently mysterious, that virtually any question we can ask has unresolved aspects. Nevertheless, we here highlight two major issues that are especially timely.

### 4.3.1  Abundances trends in exoplanets: Emerging reality, or illusion?

The abundances of heavy elements in a planet (its "metallicity") are indicative of its origin and evolution.  In our Solar System, there is an inverse correlation between metallicity and planetary mass.   The lowest mass planets of our Solar System are rocky, with little to no hydrogen, whereas our gas giants have abundant hydrogen.  We already know from mass and radius measurements of transiting exoplanets that a similar relation holds for the bulk composition of exoplanets.  Beyond bulk composition, we expect that the atmospheres and envelopes of low mass planets should be enriched in heavy elements relative to more massive planets, as a result of minimal gas accretion during the late stages of planet formation.  Moreover, differentiation within a planet (heavy elements sink) will also play a role.  In an effort to better understand planetary formation and differentiation, a major current theme is an attempt to define the mass-composition relation(s) that characterize exoplanetary atmospheres (e.g., Kreidberg et al. 2014b, Fraine et al. 2014), by using atmospheric water vapor (i.e., oxygen) as a proxy for metallicity.

Defining the relationship between exoplanetary mass and envelope metallicity will be difficult for multiple reasons.  First, the individual lines of water vapor are saturated in a transit spectrum, reducing their sensitivity to abundance.  Second, line formation for transit spectroscopy occurs over a narrow range in longitude, making it difficult to measure the temperature (e.g., from phase curves). The water absorption is sensitive to temperature via the atmospheric scale height, and also via the lower state populations of the transitions. Third, there is a degeneracy between cloud and gas abundances: a given spectrum can be produced by a high fractional abundance in a small gas column density - limited by cloud opacity, or by a low fractional abundance in a large gas column density for a clear atmosphere.  Fourth, there are chemical sinks for oxygen other than water vapor.  For example, Helling et al. (2016) assert that "no one value for the metallicity and the C/O ratio can be used to describe an extrasolar planet", due to variation in cloud composition versus height.  Finally, the intrinsic relation between exoplanet mass and metallicity is likely to be more complex than a simple inverse function.  Already, population models (Fortney et al. 2013) predict substantial scatter in metallicity at a given planetary mass, so more than a few well-measured exoplanets will be needed to define the full reality.

In spite of the above challenges, there are reasons to be optimistic about exoplanet abundances for the long term.  Several favorable factors are at play:  first, oxygen is a reasonable proxy for metallicity, because oxygen is the third most cosmically-abundant element (after hydrogen and helium).  Second, pure water vapor will not condense at the temperatures that characterize hot transiting planets, so water vapor is observable.  Third, observational techniques for water vapor are making good progress, as described in Sec. 4.2.1.

Secondary eclipse spectroscopy used in combination with transit spectroscopy will be needed to reduce uncertainties due to temperature and scale height.  The continuum-slope versus water absorption strength relation defined by Sing et al. (2016, see Figure 10) can help to break the cloud-abundance degeneracy.  Further advances in understanding cloud chemistry can help to evaluate the amount of unobservable oxygen, and large ongoing HST Cycle-23 and -24 programs will measure water vapor abundances for a statistically significant sample of planets.  Beware of illusions that will develop during this process, but we expect that those illusions will transition into reality in the long term.

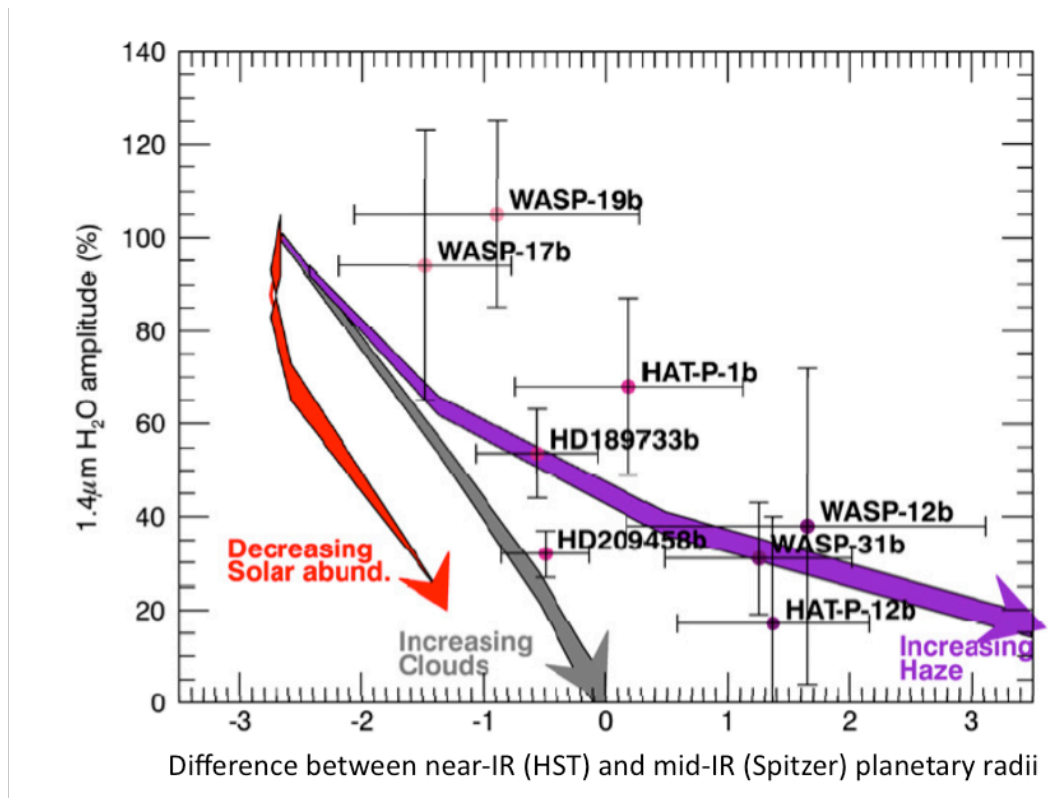

Figure 10. Water absorption (ordinate) versus IR continuum slope (abscissa) from Sing et al. (2016). The red, gray and purple tracks are model calculations of the loci of clear models with varying metallicity (red), and solar abundance models with increasing clouds (gray) or small-particle haze (purple).

### 4.3.2 Existence and nature of super-Earth atmospheres

We know from masses and radii of mini-Neptunes, extending into the super-Earth regime, that those planets have gaseous envelopes and therefore have atmospheres (see Sec. 2.1). However, for planets smaller than about 1.5 $R_\oplus$, we cannot be sure whether they have atmospheres. A current observational frontier is the attempt to detect super Earth atmospheres, either using transit spectroscopy (ongoing HST Cycle-23 and -24 programs), or measuring their albedos. A sufficiently high albedo (e.g., 0.6, see below) will imply the existence of clouds, hence of atmospheres. HST/WFC3 transit spectroscopy is sensitive to one specific band of water vapor, in one narrow wavelength range (1.1-1.7 μm), whereas JWST will sense a much larger IR wavelength range (Sec. 5.2). Meanwhile, transit spectroscopy in the optical and UV continuum could reveal the presence of a scattering atmosphere. Geometric albedos for super-Earths were inferred by Demory (2014), using secondary eclipses of the most favorable super-Earths in the Kepler data, and extrapolating to a larger sample based on a hierarchical Bayesian analysis. Demory infers a median geometric albedo between 0.06 and 0.11. Sheets and Deming (2014, 2016) use an alternate statistical procedure to infer the average albedo of Kepler's super-Earths, by directly combining secondary eclipse photometry for tens of thousands of eclipses of multiple planets with similar radii, scaling their eclipse durations to a common time scale. Their sample is restricted to close-in planets, where the reflected light signal would exceed 10 ppm for a geometric albedo of unity. Such planets will have short orbital periods, and are necessarily hot. They find that 1-2 $R_\oplus$ planets in this group have an average albedo of 0.11 +/- 0.06. That is not a secure detection of reflected light, but it does securely imply that hot super-Earths are dark. However, a few hot super-Earths, such as Kepler-10b, have much higher geometric albedos, near 0.6 (Sheets and Deming 2014).

One possibility consistent with the current observations is that all hot super-Earths have atmospheres, and a few have highly reflective clouds. Although high albedos imply clouds, hence an atmosphere, the inverse is not necessarily true: low albedos could result from either a clear atmosphere, or a dark surface (or both). So it is also possible that only the most reflective super-Earths have atmospheres. Testing these hypotheses will require powerful new observational facilities, that we now describe.

## 5.0 Future facilities to investigate exoplanetary atmospheres

Current facilities like the Hubble and Spitzer Space Telescopes will continue to provide valuable exoplanetary data for years to come, but here we focus on several future facilities that will have particularly powerful applications to exoplanetary atmospheres.

### 5.1 Ground-based Extremely Large Telescopes (ELTs)

A new generation of 30-meter class telescopes will have many applications to exoplanetary atmospheres. Adaptive optics and the large light gathering power of these telescopes will provide greatly improved spectra of young giant planets that are directly imaged. As for transiting planets, ground-based photometric precision cannot rival space-borne facilities. Nevertheless, some types of combined light spectroscopy from the ground can compete with space observatories for two reasons. First, spectra require many photons to provide enough light given the dispersion in wavelength, and the ELTs will be unsurpassed in light gathering power. Second, comparing closely adjacent wavelengths (e.g., when measuring atomic alkali lines in transit, Redfield et al. 2008) can facilitate common-mode removal of fluctuations in telluric opacity. Hence we expect that spectroscopic techniques such as Doppler deconvolution (Sec. 3.1.3) will benefit greatly from the ELTs. One far reaching, but ultimate possibility for ELT spectroscopy is detection (in tens of transits) of molecular oxygen, a biosignature, in the atmosphere of a transiting habitable exoplanet (Snellen et al. 2013).

### 5.2 The James Webb Space Telescope (JWST)

The launch of JWST in 2018 will be a boon for the study of exoplanetary atmospheres, for three reasons. First, and most important, JWST will be a spectroscopic powerhouse. JWST's four instruments provide options for many spectroscopic modes, from low to moderately high spectral resolving power. Second, it operates at IR wavelengths from just under 1 micron to beyond 20 microns. The IR is the spectral region where strong vibration-rotation bands of planetary atmospheric molecules are found. Whereas HST/WFC3 provides access to water vapor but very little sensitivity to carbon-containing molecules, JWST will cover the strong fundamental bands of methane, carbon monoxide, and carbon dioxide. Third, JWST's 6-meter aperture will increase the photon-limited signal-to-noise by a factor of 2.5 when compared to Hubble at the same wavelength.

The exoplanetary atmospheric science envisioned for JWST include high signal-to-noise observations of hot Jupiter atmospheres (Stevenson et al. 2016), atmospheric compositions of warm Neptunes and super-Earths (Greene et al. 2016, see Figure 11), and detection of the major molecules in the atmospheres of habitable super-Earths (Deming et al. 2009), and possibly the detection of biosignatures (although we will have to be lucky for that). A good review of possible exoplanetary science from JWST is Beichman et al. (2014), and more recently Cowan et al. (2015), and Greene et al. (2016).

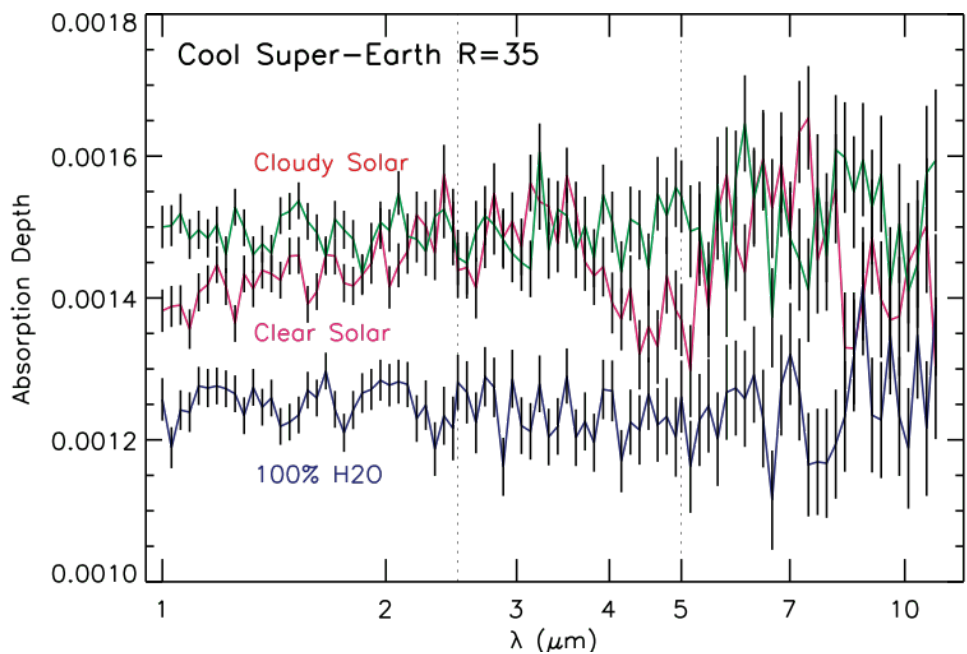

Figure 11. Simulated JWST secondary eclipse spectra of a "cool" super-Earth (T = 500K) using JWST, from Greene et al. (2016). The spectra are for a single eclipse, and for clear and cloudy solar abundance atmospheres, and a clear water vapor atmosphere. The signal-to-noise ratio will increase proportional to the square root of the number of eclipses, but already these spectra detect water absorption in the pure water atmosphere (averaged over the spectrum, not detection of each individual feature).

## 5.3 HabEx and LUVOIR

NASA's most ambitious missions currently under study[2,3] for consideration by the 2020 Decadal Survey, to address exoplanetary science are HabEx and LUVOIR (although the latter will address a wide range of science from cosmology to black holes, to exoplanets.) The HabEx Study Team is considering apertures in the 4- to 8-meter range, with a coronagraph and/or an external occulter (Spitzer 1962, and e.g., Cash 2006), to achieve extremely high contrast imaging (better than one part in 10 billion) of terrestrial planets orbiting nearby solar-type stars. LUVOIR would address similar exoplanetary science, with a larger aperture (> 10 meters is being discussed). The major advantage of these missions over current direct imaging facilities is that they would be capable of imaging truly Earth-like planets in the habitable zones of solar-type stars. They could define the frequency of occurrence of those Earth-like worlds, and - for the most favorable cases - detect the presence of oceans and continents via their rotational light curves (e.g. Ford et al., 2001 Cowan et al. 2009), and possibly biosignature gases in their spectra (Seager, Bains and Hu, 2013).

---

[2] http://www.jpl.nasa.gov/habex/

[3] http://asd.gsfc.nasa.gov/luvoir/

In this review, we have emphasized that the study of exoplanetary atmospheres often moves from illusion to reality, as the first tentative results are superceded by better or more extensive observations. We expect that advanced missions such as HabEx and LUVOIR will experience the same evolution. Even before HabEx and LUVOIR, the WFIRST coronagraph (Zhao, 2014) and potential Starshade Rendezvous Mission with WFIRST (Seager et al. 2015) will uncover the first tantalizing hints of true Earth twins. During and beyond HabEx and LUVOIR a continued flow of higher quality observations will clarify our view of distant worlds.

### Acknowledgements

We thank Dr. Nick Cowan for an insightful review of this paper, and we are grateful to Dr. Leslie Rogers for providing Figure 1, and to Dr. Hannah Wakeford for Figure 3.

# References

Adams, E. R., S. Seager, and L. Elkins-Tanton (2008), Ocean planet or thick atmosphere: on the mass-radius relationship for solid exoplanets with massive atmospheres, ApJ, 673, 1160-1164, doi:10.1086/524925.

Anglada-Escude, G., et al. (2016), A terrestrial planet candidate in a temperate orbit around Proxima Centauri, Nature, 536, 437-440, doi:10.1038/nature19106.

Atreya, S. K. and T. M. Donahue (1979), Models of the jovian upper atmosphere, Reviews of Geophysics and Space Physics, 17, 388, doi:10.1029/RG017i003p00388.

Bean, J. L., E. Miller-Ricci, and D. Homeier (2010), A ground-based transmission spectrum of the super-Earth exoplanet GJ1214b, Nature, 468, 669-672, doi:10.1038/nature09596.

Beichman, C., et al. (2014), Observations of transiting exoplanets with the James Webb Space Telescope (JWST), PASP, 126, 1134-1173, doi:10.1086/679566.

Benneke, B. (2016), Strict upper limits on the carbon-to-oxygen ratios of eight hot Jupiters from self-consistent atmospheric retrieval, submitted to ApJ, astro-ph/1504.07655.

Berdyugina, S. V., A. V. Berdyugin, D. M. Fluri, and V. Piirola, V. (2008), First detection of polarized scattered light from an exoplanetary atmosphere, ApJ, 673, L83-L86, doi:10.1086/527320.

Birkby, J. L., et al. (2013), Detection of water absorption in the day side atmosphere of HD 189733b using ground-based high-resolution spectroscopy at 3.2 microns, MNRAS, 436, L35-L39, doi:10.1093/mnrasl/slt107.

Bott, K., et al. (2016), The polarization of HD 189733, MNRAS, 459, L109-L113, doi:10.1093/mnrasl/slw046.

Brewer, J. M., D. A. Fischer, J. A. Valenti, and N. Piskunov (2016), Spectral properties of cool stars: extended abundance analysis of 1,617 planet-search stars, ApJS, 225, id.32, doi:10.3847/0067-0049/225/2/32.

Brogi, M., et al. (2012), The signature of orbital motion from the dayside of the planet τ Boötis b, Nature, 486, 502-504, doi:10.1038/nature11161.

Brogi, M., R. J. de Kok, J. L. Birkby, H. Schwarz, and I. A. G. Snellen (2014), Carbon monoxide and water vapor in the atmosphere of the non-transiting exoplanet HD 179949b, A&A, 565, id.A124, doi:10.1051/0004-6361/201423537.

Burrows, A. S. (2014), Highlights in the study of exoplanet atmospheres, Nature, 513, 345-352, doi:10.1038/nature13782.

Cash, W. (2006), Detection of Earth-like planets around nearby stars using a petal-shaped occulter, Nature, 442, 51-53, doi:10.1038/nature04930.

Charbonneau, D., R. W. Noyes, S. G. Korzennik, P. Nisenson, S. Jha, S. S. Vogt, and R. I. Kibrick (1999), An upper limit on the reflected light from the planet orbiting the star τ Bootis, ApJ, 522, L145-L148, doi:10.1086/312234.

Charbonneau, D., T. M. Brown, D. W. Latham, and M. Mayor (2000), Detection of planetary transits across a Sun-like star, ApJ, 529, L45-L48, doi:10.1086/312457.

Charbonneau, D., T. M. Brown, R. W. Noyes, and R. L. Gilliland (2002), Detection of an extrasolar planet atmosphere, Ap.J., 568, 377-384, doi:10.1086/338770.

Charbonneau, D., et al. (2005), Detection of thermal emission from an extrasolar planet, ApJ, 626, 523-529, doi:10.1086/429991.

Chilcote, J., et al. (2015), The First H-band spectrum of the giant planet β Pictoris b, ApJ, 798, L3-L7, doi:10.1088/2041-8205/798/1/L3.

Collier-Cameron, A., K. Horne, A. Penny, and D. James (1999), Probable detection of starlight reflected from the giant planet orbiting τ Boötis, Nature, 402, 751-755, doi:10.1038/45451.

Cowan, N. B., and E. Agol (2008), Inverting phase functions to map exoplanets, ApJ, 678, L129-L132, doi:10.1086/588553.

Cowan, N. B., et al. (2009), Alien maps of an ocean-bearing world, ApJ, 700, 915-923, doi:10.1088/0004-637X/700/2/915.

Cowan, N. B., and E. Agol (2011), The statistics of albedo and heat recirculation on hot exoplanets, ApJ, 729, id.54, doi:10.1088/0004-637X/729/1/54.

Cowan, N. B., et al. (2012), Thermal phase variations of WASP-12b: defying predictions, ApJ, 747, id.82, doi:10.1088/0004-637X/747/1/82.

Cowan, N. B., et al. (2015), Characterizing transiting planet atmospheres through 2025, PASP, 127, 311-327, doi:10.1086/680855.

Crossfield, I. J. M., et al. (2012), Spitzer/MIPS 24 μm observations of HD 209458b: three eclipses, two and a half transits, and a phase curve corrupted by instrumental sensitivity variations, ApJ, 752, id.81, doi:10.1088/0004-637X/752/2/81.

Crossfield, I. J. M., et al. (2014), A global cloud map of the nearest known brown dwarf, Nature, 505, 654-656, doi:10.1038/nature12955.

Crossfield, I. J. M. (2015), Observations of exoplanet atmospheres, PASP, 127, 941-960, doi:10.1086/683115.

Crouzet, N., P. R. McCullough, C. Burke, and D. Long (2012), Transmission spectroscopy of exoplanet XO-2b observed with Hubble Space Telescope NICMOS, ApJ, 761, 7-19, doi:10.1088/0004-637X/761/1/7.

Crouzet, N., P. R. McCullough, D. Deming, and N. Madhusudhan (2014), Water vapor in the spectrum of the extrasolar planet HD 189733b. II. the eclipse, ApJ, 795, id.166, doi:10.1088/0004-637X/795/2/166.

Currie, T., et al. (2014), Deep thermal infrared imaging of HR 8799 bcde: new atmospheric constraints and limits on a fifth planet, ApJ, 795, id.133, doi:10.1088/0004-637X/795/2/133.

Deming, D., S. Seager, L. J. Richardson, and J. Harrington (2005), Infrared radiation from an extrasolar planet, Nature, 434, 740-743, doi:10.1038/nature03507.

Deming, D., J. Harrington, S. Seager, and L. J. Richardson (2006), Strong infrared emission from the extrasolar planet HD 189733b, ApJ, 644, 560-564, doi:10.1086/503358.

Deming, D., et al. (2009), Discovery and characterization of transiting super-Earths using an all-sky transit survey and follow-up by the James Webb Space Telescope, PASP, 121, 952-967, doi:10.1086/605913.

Deming, D., et al. (2011), Kepler and ground-based transits of the exo-Neptune HAT-P-11b, ApJ, 740, id.33, doi:10.1088/0004-637X/740/1/33.

Deming, D., et al. (2013), Infrared transmission spectroscopy of the exoplanets HD 209458b and XO-1b using the Wide Field Camera-3 on the Hubble Space Telescope, ApJ, 774, id.95, doi:10.1088/0004-637X/774/2/95.

Demory, B. O., et al. (2011), The high albedo of the hot Jupiter Kepler-7b, ApJ, 735, L12-L15, doi:10.1088/2041-8205/735/1/L12.

Demory, B. O., et al. (2013), Inference of inhomogeneous clouds in an exoplanet atmosphere, ApJ, 776, id.L25, doi:10.1088/2041-8205/776/2/L25.

Demory, B.-O. (2014), The albedos of Kepler's close-in super-Earths, ApJ, 789, id.L20, doi:10.1088/2041-8205/789/1/L20.

Demory, B. O., et al. (2016), A map of the large day-night temperature gradient of a super-Earth exoplanet, Nature, 532, 207-209, doi:10.1038/nature17169.

Des Marais, D. J., et al. (2002), Remote sensing of planetary properties and biosignatures on extrasolar terrestrial planets, Astrobiology, 2, 153-181, doi: 10.1089/15311070260192246.m.

de Wit, J., M. Gillon, B.-O. Demory, and S. Seager (2012), Towards consistent mapping of distant worlds: secondary-eclipse scanning of the exoplanet HD 189733b, A&A, 548, id.A128, doi:10.1051/0004-6361/201219060.

de Wit, J., et al. (2016a), A combined transmission spectrum of the Earth-sized exoplanets TRAPPIST-1 b and c, Nature, 537, 69-72, doi:10.1038/nature18641.

de Wit, J., et al. (2016b), Direct measure of radiative and dynamical properties of an exoplanet atmosphere, ApJ, 820, id.L33, doi:10.3847/2041-8205/820/2/L33.

Diamond-Lowe, H., K. B. Stevenson, J. L. Bean, M. R. Line, and J. J. Fortney (2014), New analysis indicates no thermal inversion in the atmosphere of HD 209458b, ApJ, 796, 66-72, doi:10.1088/0004-637X/796/1/66.

Ehrenreich, D., and J.-M. Desert (2011), Mass-loss rates for transiting exoplanets, A&A, 529, id.A136, doi:10.1051/0004-6361/201016356.

Elkins-Tanton, L. T. and S. Seager (2008), Ranges of atmospheric mass and composition of super-Earth exoplanets, ApJ, 685, 1237-1246, doi:10.1086/591433.

Esteves, L. J., E. J. W. De Mooij, and R. Jayawardhana (2013), Changing phases of alien worlds: probing atmospheres of Kepler planets with high-precision photometry, ApJ, 804, id.150, doi:10.1088/0004-637X/804/2/150.

Evans, T. M., et al. (2016), Detection of H2O and evidence for TiO/VO in an ultra-hot exoplanet atmosphere, ApJ, 822, id.L4, doi:10.3847/2041-8205/822/1/L4.

Fortney, J. J. (2012), On the carbon-to-oxygen ratio measurement in nearby Sun-like stars: implications for planet formation and the determination of stellar abundances, ApJ, 747, id.L47, doi:10.1088/2041-8205/747/2/L27.

Fortney, J. J., et al. (2013), A framework for characterizing the atmospheres of low-mass low-density transiting planets, ApJ, 775, id.80, doi:10.1088/0004-637X/775/1/80.

Fraine, J., et al. (2014), Water vapour absorption in the clear atmosphere of a Neptune-sized exoplanet, Nature, 513, 526-529, doi:10.1038/nature13785.

Gandolfi, D., et al. (2013), Kepler-77b: a very low albedo, Saturn-mass transiting planet around a metal-rich solar-like star, A&A, 557, id.A74, doi:10.1051/0004-6361/201321901.

Garcia Munoz, A., et al. (2012), Glancing views of the Earth: from a lunar eclipse to an exoplanetary transit, ApJ, 755, id.103, doi:10.1088/0004-637X/755/2/103.

Gibson, N. P., F. Pont, and S. Aigrain (2011), A new look at NICMOS transmission spectroscopy of HD 189733, GJ-436 and XO-1: no conclusive evidence for molecular features, MNRAS, 411, 2199-2213, doi: 10.1111/j.1365-2966.2010.17837.x.

Greene, T. P., M. R. Line, C. Montero, J. J. Fortney, J. Lustig-Yaeger, and K. Luther (2016), Characterizing transiting exoplanet atmospheres with JWST, ApJ, 817, id.17, doi:10.3847/0004-637X/817/1/17.

Grillmair, C. J., et al. (2008), Strong water absorption in the dayside emission spectrum of the planet HD189733b, Nature, 456, 767-769, doi:10.1038/nature07574.

Guillot, T., A. Burrows, W. B. Hubbard, J. I. Lunine, and D. Saumon (1996), Giant planets at small orbital distances, ApJ, 459, L35-L38, doi:10.1086/309935.

Hansen, C. J., J. C. Schwartz, and N. B. Cowan (2014), Features in the broad-band eclipse spectra of exoplanets: signal or noise?, MNRAS, 444, 3632-3640, doi: 10.1093/mnras/stu1699.

Haynes, K., A. M. Mandell, N. Madhusudhan, D. Deming, and H. Knutson (2015), Spectroscopic evidence for a temperature inversion in the dayside atmosphere of hot Jupiter WASP-33b, ApJ, 806, 146-157, doi:10.1088/0004-637X/806/2/146.

Helling, Ch., et al. (2016), The mineral clouds on HD 209458b and HD 189733b, MNRAS, 460, 855-883, doi:10.1093/mnras/stw662.

Heng, K. (2016), A cloudiness index for transiting exoplanets based on the sodium and potassium lines: tentative evidence for hotter atmospheres being less cloudy at visible wavelengths, ApJ, 826, id.L16, doi:10.3847/2041-8205/826/1/L16.

Heng, K. and P. Kopplara (2012), On the Stability of Super-Earth Atmospheres, ApJ, 754, id.60, doi:10.1088/0004-637X/754/1/60.

Heng, K., and A. S. Showman (2015), Atmospheric dynamics of hot exoplanets, AREPS, 43, 509-540, doi:10.1146/annurev-earth-060614-105146.

Henry, G. W., G. W. Marcy, R. P. Butler, and S. S. Vogt (2000), A transiting "51 Peg-like" planet, ApJ, 529, L41-L44, doi:10.1086/312458.

Hui, L., and S. Seager (2002), Atmospheric lensing and oblateness effects during an extrasolar planetary transit, ApJ, 572, 540-555, doi:10.1086/340017.

Huitson, C. M., et al. (2012), Temperature-pressure profile of the hot Jupiter HD 189733b from HST sodium observations: detection of upper atmospheric heating, MNRAS, 422, 2477-2488, doi:10.1111/j.1365-2966.2012.20805.x.

Huitson, C. M., et al. (2013), An HST optical-to-near-IR transmission spectrum of the hot Jupiter WASP-19b: detection of atmospheric water and likely absence of TiO, MNRAS, 434, 3252-3274, doi:10.1093/mnras/stt1243.

Hunten, D. M. (1990), Kuiper Prize Lecture - Escape of atmospheres, ancient and modern, Icarus, 85, pp.1-20, doi:10.1016/0019-1035(90)90100-N.

Ingalls, J. G., et al. (2016), Repeatability and accuracy of exoplanet eclipse depths measured with post-cryogenic Spitzer, AJ, in press, astro-ph/1601.05101.

Ingraham, P., et al. (2014), Gemini Planet Imager spectroscopy of the HR 8799 planets c and d, ApJ, 794, id.L15, doi:10.1088/2041-8205/794/1/L15.

Jensen, A. G., et al. (2011), A survey of alkali line absorption in exoplanetary atmospheres, ApJ, 743, 203-216, doi:10.1088/0004-637X/743/2/203.

Kataria, T., et al. (2016), The atmospheric circulation of a nine-hot-Jupiter sample: probing circulation and chemistry over a wide phase space, ApJ, 821, id.9, doi: 10.3847/0004-637X/821/1/9.

Kipping, D. M., and Spiegel, D. S. (2011), Detection of visible light from the darkest world, MNRAS, 417, L88-L92, doi:10.1111/j.1745-3933.2011.01127.x.

Knutson, H. A., et al. (2007), A map of the day-night contrast of the extrasolar planet HD 189733b, Nature, 447, 183-186, doi:10.1038/nature05782.

Knutson, H. A., D. Charbonneau, L. E. Allen, A. Burrows, and S. T. Megeath (2008), The 3.6-8.0 μm broadband emission spectrum of HD 209458b: evidence for an atmospheric temperature inversion, ApJ, 673, 526-531, doi:10.1086/523894.

Knutson, H. A., et al. (2009), The 8 μm phase variation of the hot Saturn HD149026b, ApJ, 703, 769-784, doi:10.1088/0004-637X/703/1/769.

Knutson, H. A., et al. (2012), 3.6 and 4.5 μm phase curves and evidence for non-equilibrium chemistry in the atmosphere of extrasolar planet HD 189733b, ApJ, 754, id.22, doi:10.1088/0004-637X/754/1/22.

Knutson, H. A., et al. (2014), Hubble Space Telescope near-IR transmission spectroscopy of the super-Earth HD 97658b, ApJ, 794, id.155, doi: 10.1088/0004-637X/794/2/155.

Konopacky, Q. M., T. S. Barman, B. A. Macintosh, and C. Marois (2013), Detection of carbon monoxide and water absorption lines in an exoplanet atmosphere, Science, 339, 1398-1401, doi:10.1126/science.1232003.

Kreidberg, L., et al. (2014a), Clouds in the atmosphere of the super-Earth exoplanet GJ1214b, Nature, 505, 69-72, doi:10.1038/nature12888.

Kreidberg, L., et al. (2014b), A precise water abundance measurement for the Hot Jupiter WASP-43b, ApJ, 793, id.L27, doi:10.1088/2041-8205/793/2/L27.

Kreidberg, L., et al. (2015), A detection of water in the transmission spectrum of the hot Jupiter WASP-12b and implications for its atmospheric composition, ApJ, 814, 66-80, doi:10.1088/0004-637X/814/1/66.

Kuchner, M. and S. Seager (2005), Extrasolar carbon planets, astro-ph/0504214.

Lammer, H., et al. (2003), Atmospheric loss of exoplanets resulting from stellar X-Ray and extreme-ultraviolet heating, ApJ, 598, L121-L124, doi:10.1086/380815.

Laughlin, G., et al. (2009), Rapid heating of the atmosphere of an extrasolar planet, Nature, 457, 562-564, doi:10.1038/nature07649.

Lewis, N., et al. (2013), Orbital phase variations of the eccentric giant planet HAT-P-2b, ApJ, 766, id.95, doi:10.1088/0004-637X/766/2/95.

Lewis, N., et al. (2014), Atmospheric circulation of eccentric hot Jupiter HAT-P-2b, ApJ, 795, id.150, doi:10.1088/0004-637X/795/2/150.

Line, M. R., H. Knutson, A. S. Wolf, and Y. L. Yung (2014), A systematic retrieval analysis of secondary eclipse spectra. II. a uniform analysis of nine planets and their C to O ratios, ApJ, 783, 70-82, doi:10.1088/0004-637X/783/2/70.

Line, M. R. and Parmentier, V. (2016), The influence of nonuniform cloud cover on transit transmission spectra, ApJ, 820, id.78, doi: 10.3847/0004-637X/820/1/78.

Line, M. R., et al. (2016), No thermal inversion and a solar water abundance for the hot Jupiter HD209458b from HST WFC3 emission spectroscopy, AJ 152, id.203, doi:10.3847/0004-6256/152/6/203.

Lockwood, A. C., et al. (2014), Near-IR direct detection of water vapor in Tau Boötis b, ApJ, 783, id.L29, doi:10.1088/2041-8205/783/2/L29.

Lodders, K. (2004), Jupiter formed with more tar than ice, ApJ, 611, 587, doi:10.1086/421970.

Lopez, E. D., and J. J. Fortney (2014), Understanding the mass-radius relation for sub-Neptunes: radius as a proxy for composition, ApJ, 792, 1-17, doi:10.1088/0004-637X/792/1/1.

Lovis, C., et al. (2016), Atmospheric characterization of Proxima b by coupling the SPHERE high-contrast imager to the ESPRESSO spectrograph, submitted to A&A, astro-ph/1609.03082.

Machalek, P., et al. (2009), Detection of thermal emission of XO-2b: evidence for a weak temperature inversion, ApJ, 701, 514-520, doi:10.1088/0004-637X/701/1/514.

Macintosh, B., et al. (2007), Adaptive optics for direct detection of extrasolar planets: the Gemini Planet Imager, Comptes Rendus - Physique, 8, 365-373, doi:10.1016/j.crhy.2007.04.007.

Madhusudhan, N., and S. Seager (2009), A temperature and abundance retrieval method for exoplanet atmospheres, ApJ, 707, 24, doi:10.1088/0004-637X/707/1/24.

Madhusudhan, N., et al. (2011), A high C/O ratio and weak thermal inversion in the atmosphere of exoplanet WASP-12b, Nature, 469, 64-67, doi:10.1038/nature09602.

Madhusudhan, N., N. Crouzet, P. R. McCullough, D. Deming, and C. Hedges (2014), $H_2O$ abundances in the atmospheres of three hot Jupiters, ApJ, 791, id.L9, doi: 10.1088/2041-8205/791/1/L9.

Madhusudhan, N., A. Marcelino, J. I. Moses, and H. Yongyun (2016), Exoplanetary atmospheres - chemistry, formation conditions, and habitability, SSRv, 05/2016, pp. 1-64, doi:10.1007/s11214-016-0254-3.

Majeau, C., E. Agol and N. B. Cowan (2012), A Two-dimensional infrared map of the extrasolar planet HD 189733b, ApJ, 747, id.L20, doi:10.1088/2041-8205/747/2/L20.

Mandell, A. M., et al. (2013), Exoplanet transit spectroscopy using WFC3: WASP-12b, WASP-17b, and WASP-19b, ApJ, 779, id.128, doi:10.1088/0004-637X/779/2/128.

Marley, M. S., C. Gelino, D. Stephens, J. I. Lunine, and R. Freedman (1999), Reflected spectra and albedos of extrasolar giant planets. I. clear and cloudy atmospheres, ApJ, 513, 879-893, doi:10.1086/306881.

Marley, M. S., A. S. Ackerman, J. N. Cuzzi, and D. Kitzmann (2013), in Comparative Climatology of Terrestrial Planets, S. J. Mackwell, A. A. Simon-Miller, J. W. Harder, and M. A. Bullock (eds.), University of Arizona Press, Tucson, pp. 367-391, doi: 10.2458/azu_uapress_9780816530595-ch15.

Marois, C., et al. (2008), Direct imaging of multiple planets orbiting the star HR 8799, Science, 322, 1348-1352, doi:10.1126/science.1166585.

Masuda, K. (2014), Very low density planets around Kepler-51 revealed with transit timing variations and an anomaly similar to a planet-planet eclipse event, ApJ, 783, id.53, doi:10.1088/0004-637X/783/1/53.

Maxted, P. F. L., et al. (2013), Spitzer 3.6 and 4.5 μm full-orbit light curves of WASP-18, MNRAS, 428, 2645-2660, doi:10.1093/mnras/sts231.

Mayor, M. and D. Queloz (1995), A Jupiter-mass companion to a solar-type star, Nature, 378, 355, doi:10.1038/378355a0.

McCullough, P. R., N. Crouzet, D. Deming, and N. Madhusudhan (2014), Water vapor in the spectrum of the extrasolar planet HD 189733b. I. the transit, ApJ, 791, id.55, doi:10.1088/0004-637X/791/1/55.

Miller-Ricci, E., S. Seager, and D. Sasselov (2009), The atmospheric signatures of super-Earths: how to distinguish between hydrogen-rich and hydrogen-poor atmospheres, ApJ, 690, 1056-1067, doi:10.1088/0004-637X/690/2/1056.

Misra, A. K. and V. S. Meadows (2014), Discriminating between cloudy, hazy, and clear sky exoplanets using refraction, ApJ, 795, id.L14, doi:10.1088/2041-8205/795/1/L14.

Morris, B. M., Mandell, A. M., and Deming, D. (2013), Kepler's optical secondary eclipse of HAT-P-7b and probable detection of planet-induced stellar gravity darkening, ApJ, 764, id.L22, doi:10.1088/2041-8205/764/2/L22.

Nikolov, N., et al. (2014), Hubble Space Telescope hot Jupiter transmission spectral survey: a detection of Na and strong optical absorption in HAT-P-1b, MNRAS, 437, 46-66, doi:10.1093/mnras/stt1859.

Nymeyer, S., et al. (2011), Spitzer secondary eclipses of WASP-18b, ApJ, 742, 35-45, doi:10.1088/0004-637X/742/1/35.

Oberg, K. I., R. Murray-Clay, and E. A. Bergin (2011), The effects of snowlines on C/O in planetary atmospheres, ApJ, 743, L16-L20, doi:10.1088/20 41-8205/743/1/L16.

Parmentier, V., J. J. Fortney, A. P. Showman, C. Moreley, and M. S. Marley (2016), Transitions in the cloud composition of hot Jupiters, ApJ, 828, id.22, doi:10.3847/0004-637X/828/1/22.

Perez-Becker, D., and Showman, A. P. (2013), Atmospheric heat redistribution on hot Jupiters, ApJ, 776, id.134, doi:10.1088/0004-637X/776/2/134.

Rajan, A., et al. (2015), Characterizing the atmospheres of the HR8799 planets with HST/WFC3, ApJ, 809, id.L33, doi:10.1088/2041-8205/809/2/L33.

Redfield, S., M. Endl, W. D. Cochran, and L. Koesterke (2008), Sodium absorption from the exoplanetary atmosphere of HD 189733b detected in the optical transmission spectrum, ApJ, 673, id.L87, doi:10.1086/527475.

Rogers, L. A. (2015), Most 1.6 Earth-radius planets are not rocky, ApJ, 801, id.41, doi:10.1088/0004-637X/801/1/41.

Rowe, J., et al. (2008), The very low albedo of an extrasolar planet: MOST space-based photometry of HD209458, ApJ, 689, 1345-1353, doi:10.1086/591835.

Schaefer, L. and B. Fegley (2007), Outgassing of ordinary chondritic material and some of its implications for the chemistry of asteroids, planets, and satellites, Icarus, 186, 462-483, doi:10.1016/j.icarus.2006.09.002.

Schaefer, L. and B. Fegley (2009), Chemistry of silicate atmospheres of evaporating super-Earths, ApJ, 703, L113-L117, doi:10.1088/0004-637X/703/2/L113.

Schaefer, L. and B. Fegley (2010), Chemistry of atmospheres formed during accretion of the Earth and other terrestrial planets, Icarus, 208, 438-448, doi:10.1016/j.icarus.2010.01.026.

Schwartz, J. C. and N. B. Cowan (2015), Balancing the energy budget of short-period giant planets: evidence for reflective clouds and optical absorbers, MNRAS, 449, 4192-4203, doi:10.1093/mnras/stv470.

Seager, S., W. Bains, and R. Hu (2013), Biosignature gases in $H_2$-dominated atmospheres on rocky exoplanets, ApJ, 777, id.95, doi:10.1088/0004-637X/777/2/95.

Seager, S., and D. D. Sasselov (2000), Theoretical transmission spectra during extrasolar giant planet transits, ApJ, 537, 916-921, doi:10.1086/309088.

Seager, S., B. A. Whitney, and D. D. Sasselov (2000), Photometric light curves and polarization of close-in extrasolar giant planets, ApJ, 540, 504-520, doi: 10.1086/309292.

Seager, S., M. Kuchner, C. A. Hier-Majumder, and B. Militzer (2007), Mass-radius relationships for solid exoplanets, ApJ, 669, 1279-1297, doi: 10.1086/521346

Seager, S., and D. Deming (2009), On the method to infer an atmosphere on a tidally locked super-Earth exoplanet and upper limits to GJ 876d, ApJ, 703, 1884-1889, doi:10.1088/0004-637X/703/2/1884.

Seager, S., and D. Deming (2010), Exoplanet atmospheres, ARAA, 48, 631-672, doi:10.1146/annurev-astro-081309-130837.

Seager, S., and L. Hui (2002), Constraining the rotation rate of transiting extrasolar planets by oblateness measurements, ApJ, 574, 1004-1010, doi:10.1086/340994.

Seager, S., et al. (2015), The Exo-S probe class starshade mission, SPIE, 9605, id.96050W, doi:10.1117/12.2190378.

Sheets, H. A., and D. Deming (2014), Statistical eclipses of close-in Kepler sub-Saturns, ApJ, 794, id.133, doi:10.1088/0004-637X/794/2/133.

Sheets, H. A., and D. Deming (2016), Average albedos of close-in super-Earths and Neptunes from statistical analysis of long cadence Kepler secondary eclipse data, to be submitted to ApJ.

Showman, A. P., C. S. Cooper, J. J. Fortney, and M. S. Marley (2008), Atmospheric circulation of hot Jupiters: three-dimensional circulation models of HD 209458b and HD 189733b with simplified forcing, ApJ, 682, 559-576, doi: 10.1086/589325.

Sidis, O., and R. Sari (2010), Transits of transparent planets—atmospheric lensing effects, ApJ, 720, 904-911, doi:10.1088/0004-637X/720/1/904.

Sing, D. K., et al. (2012), GTC OSIRIS transiting exoplanet atmospheric survey: detection of sodium in XO-2b from differential long-slit spectroscopy, MNRAS, 426, 1663-1670, doi:10.1111/j.1365-2966.2012.21938.x.

Sing, D. K., et al. (2015), HST hot-Jupiter transmission spectral survey: detection of potassium in WASP-31b along with a cloud deck and Rayleigh scattering, MNRAS, 446, 2428-2443, doi:10.1093/mnras/stu2279.

Sing, D. K., et al. (2016), A continuum from clear to cloudy hot-Jupiter exoplanets without primordial water depletion, Nature, 529, 59-62, doi:10.1038/nature16068.

Snellen, I. A. G., R. J. de Kok, E. J. W. de Mooij, and A. Simon (2010), The orbital motion, absolute mass and high-altitude winds of exoplanet HD209458b, Nature, 465, 1049-1051, doi:10.1038/nature09111.

Snellen, I. A. G., R. J. de Kok, R. le Poole, M. Brogi, and J. Birkby (2013), Finding extraterrestrial life using ground-based high-dispersion spectroscopy, ApJ, 764, id.182, doi:10.1088/0004-637X/764/2/182.

Snellen, I. A. G., R. J. de Kok, J. L. Birkby, B. Brandl, M. Brogi, C. Keller, M. Kenworthy, H. Schwarz, and R. Stuik (2015), Combining high-dispersion spectroscopy with high contrast imaging: Probing rocky planets around our nearest neighbors, A&A, 576, id.A59,10.1051/0004-6361/201425018.

Spitzer, L. (1962), The beginnings and future of space astronomy, Am. Sci., 50, 473-484.

Stevenson, K. B., et al. (2014), Thermal structure of an exoplanet atmosphere from phase-resolved emission spectroscopy, Science, 346, 838-841, doi:10.1126/science.1256758.

Stevenson, K. B., et al. (2016), Transiting exoplanet studies and community targets for JWST's Early Release Science program, PASP, 128, 967, doi:10.1088/1538-3873/128/967/094401.

Sudarsky, D., A. Burrows, and P. Pinto (2000), Albedo and reflection spectra of extrasolar giant planets, ApJ, 538, 885-903, doi:10.1086/309160.

Swain, M. R., G. Vasisht, and G. Tinetti (2008), The presence of methane in the atmosphere of an extrasolar planet, Nature, 452, 329-331, doi: 10.1038/nature06823.

Swift, D. C., et al. (2012), Mass-radius relationships for exoplanets, ApJ, 744, id.59, doi:10.1088/0004-637X/744/1/59.

Tinetti, G., et al. (2010), Probing the terminator region atmosphere of the hot-Jupiter XO-1b with transmission spectroscopy, ApJ, 712, L139-L142, doi: 10.1088/2041-8205/712/2/L139.

Todorov, K. O., D. Deming, A. Burrows, and C. J. Grillmair (2014), Updated Spitzer emission spectroscopy of the bright transiting hot Jupiter HD189733b, doi: 10.1088/0004-637X/796/2/100.

Todorov, K. O., et al. (2016), The Water abundance of the directly imaged substellar companion κ And b retrieved from a near infrared spectrum, ApJ, 823, id.14, doi: 10.3847/0004-637X/823/1/14.

Triaud, A. H. M. J. (2014), Colour-magnitude diagrams of transiting exoplanets - I. systems with parallaxes, MNRAS, 439, L61-L64, doi:10.1093/mnrasl/slt180.

Vidal-Madjar, A. (2003), An extended upper atmosphere around the extrasolar planet HD209458b, Nature, 422, 143-146, doi:10.1038/nature01448.

Vidal-Madjar, A., et al. (2011), The upper atmosphere of the exoplanet HD 209458b revealed by the sodium D lines. Temperature-pressure profile, ionization layer, and thermosphere, A&A, 527, id.A110, doi:10.1051/0004-6361/201015698.

Wakeford, H. R., et al. (2013), HST hot Jupiter transmission spectral survey: detection of water in HAT-P-1b from WFC3 near-IR spatial scan observations, MNRAS, 435, 3481-3493, doi:10.1093/mnras/stt1536.

Wakeford, H. R., et al. (2016a), Marginalizing instrument systematics in HST WFC3 transit light curves, ApJ, 819, id.10, doi:10.3847/0004-637X/819/1/10.

Wakeford, H. R., et al. (2016b), HAT-P-26b: A Neptune-mass exoplanet with primordial solar heavy element abundance," submitted to Science.

Waldmann, I. P., et al. (2013), Blind extraction of an exoplanetary spectrum through independent component analysis, ApJ, 766, 6-14, doi:10.1088/0004-637X/766/1/77.

Wiedemann, G., D. Deming and G. Bjoraker (2001), A sensitive search for methane in the infrared spectrum of τ Bootis, ApJ, 546, 1068-1074, doi:10.1086/318316.

Wiktorowicz, S. J. (2009), Nondetection of polarized, scattered light from the HD 189733b hot Jupiter, ApJ, 696, 1116-1124, doi:10.1088/0004-637X/696/2/1116.

Wilson, P. A., et al. (2015), GTC OSIRIS transiting exoplanet atmospheric survey: detection of potassium in HAT-P-1b from narrow-band spectrophotometry, MNRAS, 450, 192-200, doi:10.1093/mnras/stv642.

Wong, I., et al. (2015), 3.6 and 4.5 μm phase curves of the highly irradiated eccentric hot Jupiter WASP-14b, ApJ, 811, id.122, doi:10.1088/0004-637X/811/2/122.

Wong, I., et al. (2016), 3.6 and 4.5 μm Spitzer phase curves of the highly irradiated hot Jupiters WASP-19b and HAT-P-7b, ApJ, 823, id.122, doi:10.3847/0004-637X/823/2/122.

Wordsworth, R. (2015), Atmospheric heat redistribution and collapse on tidally locked rocky planets, ApJ, 806, id.80, doi:10.1088/0004-637X/806/2/180.

Zellem, R. T., et al. (2014), The 4.5 μm full-orbit phase curve of the hot Jupiter HD 209458b, ApJ, 790, id.53, doi:10.1088/0004-637X/790/1/53.

Zhao, F. (2014), WFIRST-AFTA coronagraph instrument overview, SPIE, 9143, id. 91430O7, doi:10.1117/12.2060319.

Zhou, Y., D. Apai, G. H. Schneider, M. S. Marley, and A. P. Showman (2016), Discovery of rotational modulations in the planetary-mass companion 2M1207b: intermediate rotation period and heterogeneous clouds in a low gravity atmosphere, ApJ, 818, id.176, doi:10.3847/0004-637X/818/2/176.